\documentclass[preprint]{aa}
\usepackage{txfonts}
\usepackage{graphicx,subfigure}
\usepackage{epsfig}
\usepackage{epsf}
\usepackage{lscape}
\usepackage{natbib}
\usepackage{rotating}
\usepackage{tabularx}
\usepackage{lscape}
\usepackage{supertabular}
\newcommand{\mic}{\,{\rm \mu m} }
\newcommand{\CaseOne}{\rm regular}    
\newcommand{\CaseTwo}{\rm colder CO}    
\newcommand{\CaseThree}{\rm warmer CO}    
\newcommand{\Archeops}{\rm Archeops$\,$}    
\newcommand{\XCO}{\rm X_{CO}} 
\newcommand{\XCOUNIT}{\rm \,H_{2}/cm^{2}/(Kkm/s)} 

\begin{document}
\title{Dust emissivity variations in the Milky Way}

\author{D. Paradis \inst{1,2,3} \and J.-Ph. Bernard \inst{2,3} \and C. M\'eny \inst{2,3}}

\institute{Spitzer Science Center, California Institute of Technology, Pasadena, CA 91125 \\
email: paradis@ipac.caltech.edu 
\and
Universit\'e de Toulouse; UPS; CESR; 9 av. du Colonel Roche, F-31028 Toulouse cedex 9, France \\
\and 
CNRS; UMR5187; F-31028 Toulouse, France\\}

\date{}
\abstract
{}
{Dust properties appear to vary according to the environment in
which the dust evolves. Previous observational indications of these
variations in the far-infrared (FIR) and submillimeter (submm)
spectral range are scarce and limited to specific regions of the
sky. To determine whether these results can be generalised to
larger scales, we study the evolution in dust emissivities from the
FIR to millimeter (mm) wavelengths, in the atomic and molecular
interstellar medium (ISM), along the Galactic plane towards the outer
Galaxy.}  {We correlate the dust FIR to mm emission
with the HI and CO emission, which are taken to trace the atomic and molecular phases,
respectively. The study is carried out using the DIRBE data from 100 to
240 $\mic$, the Archeops data from 550 $\mic$ to 2.1 mm, and the WMAP
data at 3.2 mm (W band), in regions with Galactic latitude $\rm |b|
\le 30\degr$, over the Galactic longitude range ($\rm 75\degr < l <
198\degr$) observed with Archeops. }
{In all regions studied, the emissivity spectra in both the atomic
and molecular phases are steeper in the FIR ($\rm
\beta=2.4$) than in the submm and mm ($\beta=1.5$). We find significant variations in the spectral shape of the dust emissivity as a function
of the dust temperature in the molecular phase. Regions of similar
dust temperature in the molecular and atomic gas exhibit similar
emissivity spectra. Regions where the dust is significantly colder in
the molecular phase show a significant increase in emissivity for
the range 100 - 550 $\mic$. This result supports the hypothesis of grain coagulation in
these regions, confirming results obtained over small fractions of the
sky in previous studies and allowing us to expand these results to the
cold molecular environments in general of the outer MW. We note that it is the first
time that these effects have been demonstrated by direct measurement of the
emissivity, while previous studies were based only on thermal
arguments.}
{}


\maketitle
\section{Introduction}

Measuring the dust emissivity is important to inferring the nature of
dust from its emission and also determining the dust heating
from its observed temperature. In addition, variations in the dust
emissivity may seriously affect the mass estimates inferred from
FIR to mm observations.  Dust emissivity and its possible variations
with wavelength or temperature are also critical for separating astrophysical foreground emission from the
cosmic microwave background (CMB). However, deriving dust
emissivity is generally a difficult task because it relies on both
estimating the dust temperature and the gas column density associated
with the emitting dust. Big dust Grains \citep[BG, see][for a
description of each dust component]{Desert90} are in thermal
equilibrium with the interstellar radiation field (ISRF). Their
emission is close to that of a gray-body with an equilibrium temperature near
17.5 K in the diffuse ISM \citep{Boulanger96,Lagache98}, with a
maximum in the far-infrared. The emission spectrum, assuming a fixed
dust abundance and a single grain size, follows
\begin{equation}
I_{\nu}(\lambda)=\epsilon_{\nu}(\lambda)B_{\nu}(\lambda,T_d)N_H,
\label{eq:Inu}
\end{equation}
where $\rm I_{\nu}(\lambda)$ is the specific intensity or brightness
(energy flux per unit frequency and solid angle), $\rm B_{\nu}$ is the
Planck function, T$\rm _d$ is the equilibrium temperature of grains,
$\rm N_H$ is the total column density (number of hydrogen atoms per
unit area), and $\rm \epsilon_{\nu}(\lambda)$ is the wavelength-dependent dust grain emissivity.

Although, in principle, the emissivity may depend on wavelength and
temperature \citep[e.g.,][]{Meny07}, it has been customary to assume no
temperature dependence and a power law distribution with frequency,
\begin{equation}
\epsilon_{\nu}(\lambda)=X_d \epsilon_0 \left(\frac{\lambda}{\lambda_0}
\right)^{-\beta}
\label{eq:epsilon}
\end{equation}
where $\rm X_d$ is the dust to gas mass ratio,
$\rm \epsilon_0$ is the emissivity at wavelength $\rm \lambda_0$, and
$\rm \beta$ is the emissivity spectral index. The spectral index is
usually assumed to equal 2, which is correct for crystalline grains. However,
dust grains are known to be amorphous in the diffuse ISM
\citep{Kemper04} and this could have an impact on the emissivity. Grains
are probably affected by different physical processes depending on
the environment. Their properties are likely to vary according to the
density of the medium. Different studies have highlighted changes in the
grain properties in the dense and cold medium, such as the depletion
of chemical elements \citep{Caselli99,Bacmann02}, that imply that
condensation of gaseous species is occuring on the grain surface.

A decrease in the IRAS I$\rm _{60}$/I$\rm _{100}$ ratio in several
isolated molecular clouds has also been noted
\citep[e.g.,][]{Laureijs91,Abergel94,Abergel95}, whereas this ratio
seems to be constant in diffuse regions. This could be interpreted as
a decrease of very small grain (VSG) abundance with respect to that
of the BGs in dense molecular clouds. \citet{Bernard99} analysed the
FIR dust emission from a molecular cirrus in Polaris, using the PRONAOS
balloon-borne experiment data and found a larger decrease in the BG
equilibrium temperature than predicted from the decrease in the ISRF
intensity in the cloud. The same region showed an obvious deficit in 60 $\mic$ emission. They attributed the change in the particular dust
properties of dense environments.

\citet{Stepnik03} attempted to explain the origin of the decrease in
VSG abundance and the drop in the BG temperature in dense
environments. They measured a BG temperature of 16.8 K outside a
filament in Taurus, and of 12 K at the centre of the filament, using
the PRONAOS and the IRAS data. They showed that a model that assumed the standard
properties of dust, without any spatial variations in the grain
properties, was unable to reproduce the observations. The
attenuation of the radiation field in the cloud was also unable to explain the
shape of the observed emission profiles. They proposed that the
aggregation of large dust grains and most of the VSGs in fractal
assemblies would explain both the observed 60 $\mic$ emission
deficit and the unusually cold BG temperature. They showed that
aggregation would lead to an increase in the BG emissivities by a
factor of 3 - 4, and that between 80 and 100$\%$ of the VSGs should participate in
the aggregation process.  The best-fit model showed that the VSG and BG
aggregation occurs at an extinction Av $\rm >$ 2.1 mag and at
densities higher than n$\rm _H$=3 H/cm$^3$. However, the emissivity
increase in dense environments remains indirect evidence of change in the emissivity, since
$\rm T_d$ is lower than expected for the corresponding ISRF. This
effect has never been illustrated by a direct
determination of the dust emissivity, mainly because both the dust
temperature and $\rm N_H$ are difficult to measure.

\citet{Cambresy01} analysed the DIRBE emission of the Polaris complex
over 20 deg2, to determine the dust
properties. They decomposed the emission into a warm and a cold
component constrained using the I$\rm _{60}$/I$\rm _{100}$ IRAS
intensity ratio. They determined the extinction map of this complex in
the visible, by using star counts in the V band
\citep{Cambresy99}. They calculated the submillimeter emissivity of
each component, normalised to the visible extinction, and found
that the emissivity is 4 times higher in the cold region of the complex than in the warm region. This result matches the emissivity increase
derived to explain the low dust temperature by \citet{Stepnik03}, and was based on the determination of the submm to
visible dust opacity ratio, which traces the optical properties
of the grain material, independent of the gas column density.

Two hypotheses have been proposed to explain the observations: the
formation of ice mantles \citep{Laureijs91,Laureijs96} and/or grain
aggregation \citep{Draine85,Tielens89}. The 60 $\rm \mic$ emission
deficit in dense regions can be attributed to a large
majority of VSGs participating in the aggregation process, and the VSGs
are almost totally accreted onto the aggregates, in the inner regions of
the cloud \citep{Stepnik03}. The VSGs are then connected to the
aggregate, which acts as a thermostat. Their temperature then stops
fluctuating and their contribution to the mid-IR greatly decreases, which
explains their weak IRAS 60 $\rm \mic$ emission.  Calculations, for
instance using the discrete dipole approximation (DDA) method, of
optical properties of aggregates \citep{Bazell90} show that BG
aggregates are more efficient emitters than a collection of individual grains. This increased emissivity leads to more efficient
cooling for a given ISRF intensity and therefore explains the lower
temperature observed. The calculations indicate that the aggregate
emissivity should progressively deviate from that of the individual
grains in the IR (around 20 $\rm \mic$), and remain parallel to it at
longer wavelengths, in particular over the entire submillimeter and
millimeter range.

The adjunction of VSGs in the aggregate seems to have little effect on
FIR/mm properties and simply provides a slightly higher absorptivity
of the aggregate in the near-IR \citep{Stepnik_these}. Calculations by
\citet{Stepnik03} show that a total number of at least 20 individual
BG in each aggregate is necessary to provide the required emissivity
increase. However, for the aggregation process to be efficient, the individual grains should probably be covered with an
ice mantle, to increase the sticking efficiency.

The parameter $\rm \epsilon_{\nu}$ in Eq.\,\ref{eq:Inu} is consistent with the
definition used in \cite{Boulanger96}. The emissivity $\rm \epsilon_{\nu}$
is related to the absorption efficiency $\rm Q_{abs}$, which is directly
related to the refractive index of the material composing the grain by
\begin{equation}
\epsilon_{\nu}(\lambda)=\pi a^2 Q_{abs}(\lambda,a) X_d,
\label{eq:Qabs}
\end{equation}
where a is the grain radius. Therefore, for a single grain size,
the wavelength dependence of $\rm \epsilon_{\nu}$ and $\rm Q_{abs}$ are the
same. The purpose of this paper is to determine the spectral
dependence of the dust absorption cross section in the atomic and
molecular phases of the ISM and to search systematically for evidence
of emissivity changes in the dense molecular medium, to
check whether the results obtained for a limited number of isolated clouds
can be generalised. In the following, we use Eq.\,\ref{eq:Inu} to
derive the wavelength dependence of the dust emissivity, by simply
dividing the observed emission $\rm I_{\nu}(\lambda)$ by
$\rm B_{\nu}(\lambda,T_d)N_H$, for a single temperature $\rm T_d$ for each
ISM phase. However, we note that, for a realistic grain population,
Eq.\,\ref{eq:Inu} must be integrated over the grain size
distribution. Since the dust temperature depends on the dust grain
size for a given radiation field, the black-body term in
Eq.\,\ref{eq:Inu} cannot be factorised out from the
integral. Similarly, most lines of sight (LOS) must exhibit variations in the
radiation field, leading to a mixture of dust temperature along the
LOS. As a consequence, Eq.\,\ref{eq:Inu} only measures an
``effective emissivity'', whose wavelength dependence may differ
from that of the true emissivity of the material constituent of the grain.
In Sect.\,\ref{sec_bias}, we show however that our approach is
valid over the range of dust temperature and wavelengths considered
here, and that the effective emissivity precisely recovers the
wavelength dependence of the true material emissivity in the
presence of both a realistic grain size distribution and a realistic
mixture of the radiation field along the LOS. In the following, we
therefore simply refer to $\rm \epsilon_{\nu}$ as the dust emissivity, and
use Eq.\,\ref{eq:Inu} to derive the emissivity directly.

In Sect.\,\ref{sec:data}, we present the data that we used for our study and in Sect.\,\ref{sec:convfact} we define the emission to column density conversion factors used. In
Sect.\,\ref{sec:corr}, we describe the correlation procedure between
dust emission and gas tracers. In Sect.\,\ref{sec:emissivity}, we
explain how we compute the grain emissivity. The results of this study
and the verification of our hypothesis of a single temperature along
the LOS are presented in Sects.\,\ref{sec:results} and
\ref{sec_bias}. Sections\,\ref{sec:discussion} and
\ref{sec:conclusions} are devoted to our discussion and conclusions.

\section{Data}
\label{sec:data}

\subsection{FIR data}

The Diffuse Infrared Background Experiment (DIRBE) was an infrared
photometer onboard the COBE satellite (launched in 1989) to
measure the diffuse infrared and microwave radiation from the early
universe. It observed the entire sky at 10 different wavelengths between 1 and
240 $\rm \mic$, with a 40$\rm ^{\prime}$ instantaneous angular
resolution. However, the asymmetric beam of the DIRBE instrument
convolved with the spinning of the instrument produced an effective
beam of $\rm \simeq 1\degr$ in the yearly averaged products
\citep{Cambresy01b}. For our study, we only consider data in the far-infrared at 100, 140, and 240 $\rm \mic$, since emission at shorter
wavelengths contains a large contribution from thermally
fluctuating VSGs and polycyclic aromatic hydrocarbons (PAH). The DIRBE
data at these wavelengths were calibrated with models for Uranus
(100 $\mic$) and Jupiter (140 and 240 $\mic$) together with in-flight
beam shape measurements. This calibration matched that of the
FIRAS instrument, which incorporated an absolute
calibrator, within a small correction factor, which we did not
apply in this study since it is smaller than our uncertainties \citep{Fixsen97}.\\

\Archeops was a balloon-borne experiment dedicated to the measure of
the temperature fluctuations in the CMB. The focal plane instrument, a
multi-band photometer, worked in four bands centered on 550 $\rm
\mic$, 850 $\rm \mic$, 1.4 mm, and 2.1 mm. \Archeops had an angular resolution of 8$\rm ^{\prime}$ \citep[see][for a full
description of the instrument]{Benoit02}. We use data obtained during
the last flight of the instrument from Kiruna in February 2002, which
covered about 30$\rm \%$ of the sky. We note that we limit our analysis
to the region surveyed by Archeops, which is given by the range of Galactic
longitudes $\rm 75\degr \le l \le 198\degr$, in the second and third Galactic quadrants. The Archeops data were
calibrated against the FIRAS maps, at 550 $\rm \mic$ and 850 $\rm
\mic$, and with respect to the CMB dipole at 1.4 mm and 2.1 mm. We note that the calibration of the Archeops data should be accurate at
all wavelengths, since it relies on the FIRAS maps at high
frequencies, which are absolutely calibrated to $3\%$ \citep{Fixsen94},
and on the CMB dipole at low frequencies, whose brightness is
accurately known. In addition, the Archeops cosmology results derived
from the low frequency channels have been shown to be consistent with
those of the WMAP satellite at the $7\%$ accuracy \citep{Tristram05}.
Details of the \Archeops data processing are given in
\citet{Macias07}. We note that the Archeops channel at 550 $\mic$
has residual stripes, because only one detector was
available at that frequency.\\

The Wilkinson Microwave Anisotropy Probe (WMAP) measured the emission
over the entire sky in the microwave range (3.2 - 13 mm), and provided
accurate maps of the CMB fluctuations. The WMAP data, with a
13$\rm ^{\prime}$ beam, is currently the instrument of the highest angular resolution
covering the entire sky in this wavelength domain. Here, we use only the W band data (3.2 mm), because the other bands are often dominated by gas emission, such as free-free and synchrotron, or
by the anomalous foreground emission
\citep[see][]{Draine98,Bennett03}, presumably because of small spinning
particles. The calibration of the WMAP data was completed using the CMB
dipole amplitude \citep[see][]{Hinshaw03}.

\subsection{Gas tracers}

We use the Leiden/Dwingeloo survey by \citet{HartmannBurton96} for the HI data, with
sky observations above -30$\rm \degr$ of Galactic latitude, obtained
with the 25 m Dwingeloo telescope, whose angular resolution is 36$\rm
^{\prime}$ \citep[see][for more details]{Hartmann94}.

We use $\rm ^{12}CO(J=1-0)$ data compiled by \citet{Dame01}. These observations were obtained
along the Galactic plane in a Galactic latitude range from 4$\rm
\degr$ to 10$\rm \degr$, with an angular resolution of 7.5$\rm
^{\prime}$, and some observations of large clouds at higher latitude,
with an angular resolution of 15$\rm ^{\prime}$. The spatial coverage
is about 45$\rm \%$ of the sky.\\ \\

All data were projected onto the HEALPix pixelisation scheme
(Hierarchical Equal Area isoLatitude Pixelisation)\footnote{HEALPix
pixelisation distributes $\rm 12 \times Nside^2$ points as uniformly
as possible on a sphere surface, knowing that these points are divided
into (4Nside-1) parallel on latitudes and are fairly spaced on
longitudes on each one of these parallels. See
http://healpix.jpl.nasa.gov/.} with a nside=128, corresponding to
a pixel size of 0.45$\rm \degr$. For the \Archeops and WMAP first
release data, we use the published maps, and for HI and CO, we use the maps
available on the WMAP lambda web site \footnote{(http://lambda.gsfc.nasa.gov/)}. For the DIRBE data, we used our own
resampling method developed in the context of the ancillary data
for the Planck mission, computing the pixel intersection between the
original DIRBE maps in the sixcube format and the HEALPix
pixelisation. The method
preserves the photometry without significantly affecting the angular
resolution. Since the Archeops data were filtered to subtract
slow drifts, we simulated fake data stream with all data described in
this section, according to the actual Archeops scanning
strategy. Those data time-lines were treated in a similar way to the Archeops
ones, in particular by including the low frequency filtering applied to
the Archeops data, and were then reprojected into sky maps. Using the
standard HEALPix tools, all the data were corrected to the DIRBE
angular resolution (1$\rm \degr$), by convolution with a Gaussian
kernel of an appropriate size in relation to the original resolution of each
dataset.

\section{Conversion factors}
\label{sec:convfact}
Assuming that the gas is optically thin, the hydrogen column
density can be deduced from the integrated intensity of the HI
emission at 21 cm (W$\rm _{HI}$) using the relation 
\begin{equation}
N_{HI}=X_{HI}W_{HI},
\end{equation}
where $\rm X_{HI}$ is the HI integrated intensity to column density conversion factor, assumed to be equal to
$\rm 1.82\times 10^{18} \XCOUNIT$ \citep{Spitzer78}. The CO to molecular
conversion factor is given by $\rm X_{CO}=\frac{N_{H_2}}{W_{CO}}$, where $\rm
W_{CO}$ is the integrated CO emission, and we assume a fixed value of $\rm
X_{CO}$, which is defined to be the standard Galactic value proposed by
\citet{Strong88}
\begin{equation}
X_{CO}=2.3\times 10^{20} \XCOUNIT.
\end{equation}
This value was inferred from $\rm \gamma$-ray
measurements over the entire Galactic plane excluding the
Galactic center region.

\section{Correlation between IR and gas tracers}
\label{sec:corr}

We considered only regions along the Galactic plane with $\rm |b| \le
30\degr$ and pixels with sufficient HI or CO emission
($\rm W_{HI}>1000\,K/km/s$ and $\rm W_{CO}>0.5\,K/km/s$). To assess the possible
contribution from free-free,
synchrotron, and spinning dust emission on the IR emission in the WMAP W
band, we used the Planck Sky Model\footnote{see
http://www.apc.univ-paris7.fr/APC\_CS/Recherche/Adamis/\\PSM/psky.php}
\citep{Delabrouille09}.
The model indicates that the free-free emission is the main
contribution in this band and generates about 30$\%$ of the IR
emission. Its contribution was subtracted from the W band
emission, using the free-free map predicted by the Planck Sky Model. Other
contaminations (synchrotron, spinning dust) are negligible
in this band ($<$ 3$\%$), according to the same
model. The contributions from free-free, synchrotron, and spinning dust
are insignificant in other data used in this study.

We performed correlations between the infrared emission and gas
tracers, in a set of rectangular regions covering the area of
interest. Individual regions have sizes $\rm \Delta l \times \Delta
b$= 6$\rm \degr$ $\times$ 4$\rm \degr$, distributed across a regular grid
of Galactic coordinates every 3$\rm \degr$ in longitude and 2$\rm
\degr$ in latitude. In each region and each map, we subtracted a
background value computed as the median over a common background area,
defined to be the faintest half of the HI data. In this way, we ensured
that the correlation produces a null IR emission for a null column
density. This step also removes any possible residual contribution
from both the Zodiacal light and the cosmic infrared background. We
then determined the best-fit linear correlation between the FIR
emission and gas tracers using 
\begin{equation}
I_{\nu}(\lambda)=\frac{a_\nu(\lambda)}{X_{HI}} N_H ^{HI}+\frac{b_\nu(\lambda)} {2X_{CO}} N_H ^{CO}+ c_{\nu}(\lambda),
\label{eq:correl}
\end{equation}
where $\rm I_{\nu}$ is the IR emission brightness at wavelength $\rm
\lambda$, $\rm N_H ^{HI}$ and $\rm N_H ^{CO}$ are the hydrogen column
density in the atomic and molecular phases, respectively, and $\rm c_{\nu}$
is a constant. The constant term is usually small but could
account for the contribution of an additional gas component
unaccounted for in our description, such as diffuse ionized gas,
provided it is partially decorrelated from the atomic and molecular
gas distribution. We performed a $\rm \chi^2$ minimization searching
for $\rm a_\nu$ and $\rm b_\nu$ at each IR wavelength and for each
region using the interactive data language (IDL) linear regression
function ``regress''.

In a second step, we rejected regions of smaller than 15 pixels to maintain a sufficient number of pixels to perform the
correlations. The number of pixels in each region is given in tables in the Appendix. With this selection, we removed 15$\%$ of the regions. In
a third step, we removed regions with clearly unphysical values of the
dust temperatures ($>$ 1000 K). For temperatures inferred from
the spectral shape of a$\rm _{\nu}$ and b$\rm _{\nu}$ between 100 and
550 $\rm \mic$ (see Sect. \ref{sec:emissivity} for more explanation),
these unphysical values imply that at least one of the correlation
coefficients a$\rm _{\nu}$ or b$\rm _{\nu}$ in this wavelength range
is still affected by uncorrected instrumental effects, mostly residual
stripes in the Archeops 550 $\mic$ channel. This selection
removed 27$\%$ of the remaining regions. 

\section{Grain emissivity determination}
\label{sec:emissivity}

\subsection{Method}
\label{sec:method}
Using Eqs.\,\ref{eq:Inu} and \ref{eq:correl}, and assuming a single
temperature of dust in the atomic ($\rm T_d^{HI}$) and the molecular
($\rm T_d^{CO}$) phases, the emissivity of the dust associated with each
phase can be written
\begin{equation}
\label{eq:emhi}
\epsilon _{\nu}^{HI}(\lambda)=\frac{a_{\nu}(\lambda)}{B_{\nu}(T_d^{HI})}\frac{1}{X_{HI}},
\end{equation}
and
\begin{equation}
\label{eq:emco}
\epsilon_{\nu}^{CO}(\lambda)=\frac{b_{\nu}(\lambda)}{B_{\nu}(T_d^{CO})}\frac{1}{2X_{CO}},
\end{equation}
respectively. The values of the dust temperatures in the two phases
can be obtained by fitting the SED of the correlation
coefficients $\rm a_\nu$ and $\rm b_\nu$, which is equivalent, for a given
column density, to the emission spectrum of dust associated with the
atomic and molecular phases, respectively. We fitted the above SEDs
 by searching simultaneously for the best-fit average temperature and dust
emissivity index for each phase, using the correlation coefficients at
100, 140, 240, and 550 $\mic$.
Uncertainties in the emissivity values of each region are computed
using
\begin{equation}
\Delta \epsilon_{\nu}/\epsilon_{\nu}=\frac{\Delta
B_{\nu}(T)}{B_{\nu}(T)}+\frac{\Delta I_{\nu}}{I_{\nu}},
\end{equation}
where the uncertainties on the intensity for the atomic and molecular
phase are given by $\rm \Delta I_{\nu,HI}= \frac{\Delta
a_{\nu}}{X_{HI}}$ and $\rm \Delta I_{\nu,CO}= \frac{\Delta
b_{\nu}}{2X_{CO}}$, respectively. Errors in the a$\rm _{\nu}$ and b$\rm _{\nu}$
parameters correspond to the 1-$\rm \sigma$ standard deviation computed
over all valid pixels of the region. Given the Planck function 
\begin{equation}
B_{\nu}(T)=\frac{2h\nu^3}{c^2} \frac{1}{e^{\frac{h\nu}{kT}}-1},
\end{equation}
the relative error in $\rm B_{\nu}(T)$ ($\rm \Delta B_{\nu} (T)$) is related to the
error in T ($\rm \Delta T$) by 
\begin{equation}
\frac{\Delta B_{\nu} (T)}{B_{\nu}(T)}=\frac{\frac{h\nu}{kT}e^{\frac{h\nu}{kT}}}{e^{\frac{h\nu}{kT}}-1} \frac{\Delta T}{T}.
\end{equation}
Errors in temperature used to derive the error in the Planck function
are determined using a bootstrap method, when fitting for the
temperature and spectral index. We note that the temperatures are
mostly constrained by values close to the peak of the emission (i.e., 100,
140, and 240 $\mic$). However, uncertainties in the derived temperature
will strongly affect the emissivity values close the peak, because we
divide by the black-body function, a non-linear function of
the temperature near the peak. Therefore, the uncertainty in the
temperature is the main source of uncertainty in this study.
Overestimating the dust temperature by 2 K leads to an overestimate of $\rm B_{\nu}$ and therefore an underestimate of the dust emissivity by
as much as 150$\%$ and $70\%$ at 100 $\rm \mic$ and 240 $\rm \mic$,
respectively, for an average dust temperature of 16 K. For
$\rm \Delta T=0.5\,K$, these figures become $65\%$ and $35\%$, respectively.

All data used in this study follow a given flux convention (e.g., $\rm
\nu I_{\nu}=constant$ for DIRBE data), which allows us to compute unambiguously the total power received in the instrument
photometric band, regardless of the energy distribution in the
band. However, the true brightness at the reference frequency depends
on the emission spectral shape, which is unknown until a fit using a model has
converged. We therefore corrected the emissivity derived above by
dividing by the appropriate color correction factor, computed using
the filter transmission and flux convention of each instrument
used. This was done in an iterative way, starting from color
correction factors derived from a dust model, and iterating for the
true shape of each SED, until convergence was reached.

\subsection{Classification according to dust temperature}
\label{sec:cases}

\begin{figure}[!t]
\begin{center}
\includegraphics[width=8.7cm]{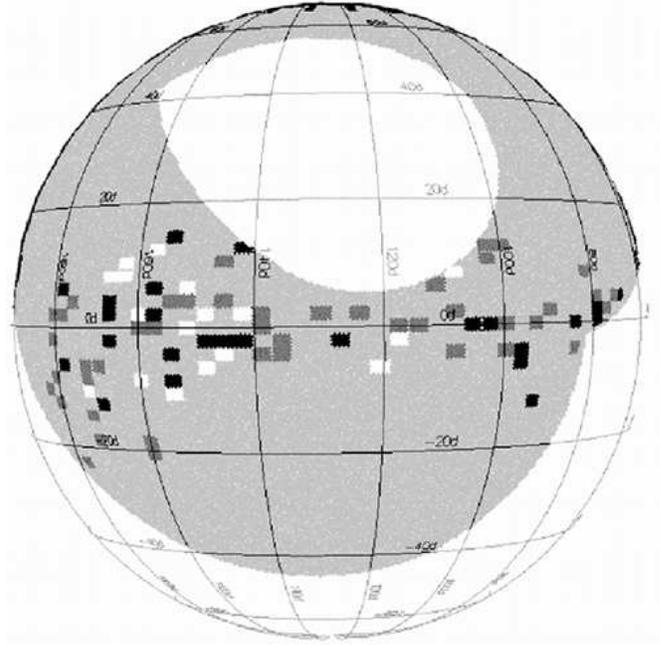}
\caption{Studied regions for each temperature case: case 1 (\CaseOne)
in white, case 2 (\CaseTwo) in grey and case 3 (\CaseThree) in black. The sky region covered by the Archeops data is also delineated.
The coordinate grid shown is in Galactic coordinates. The
map shows half the sky and is centered roughly towards the Galactic
anticenter.
\label{fig_regions}}
\end{center}
\end{figure}

\begin{figure}
\begin{center}
\includegraphics[width=8.7cm]{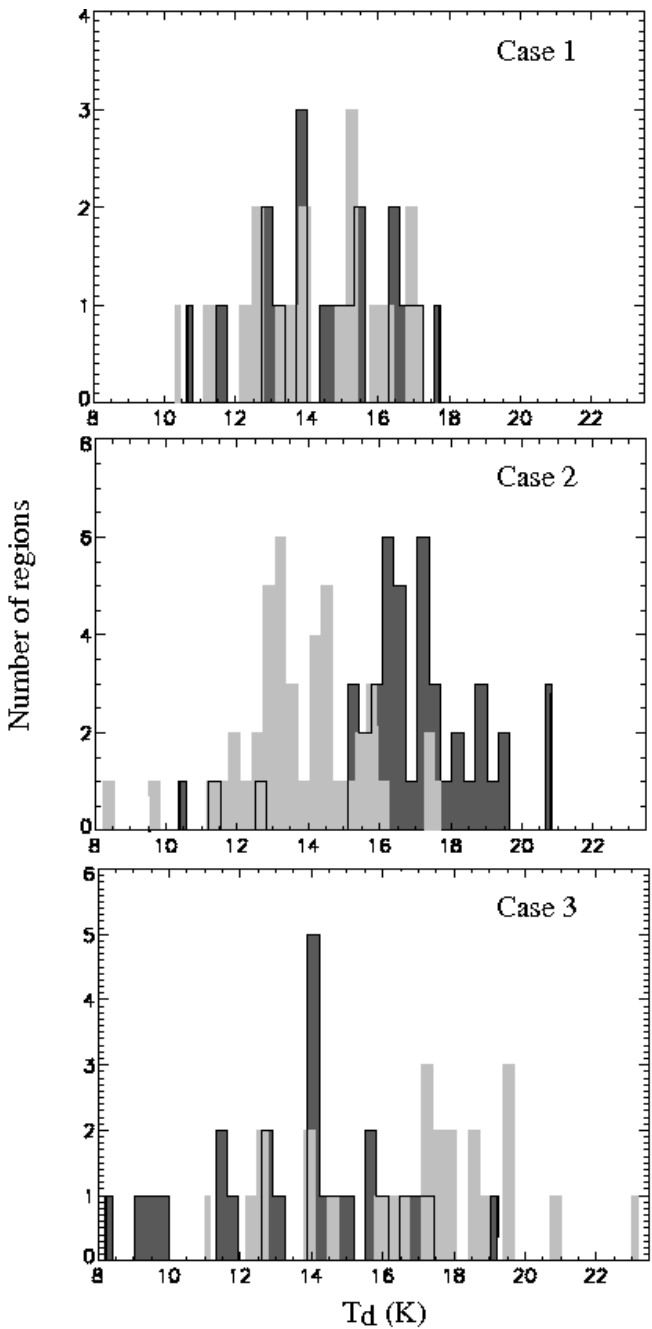}
\caption{Dust temperature histograms for each selected case: the atomic
and molecular phases are shown in dark grey with black contours and
light grey, respectively.\label{fig_hist}}
\end{center}
\end{figure}

\begin{figure}[!t]
\begin{center}
\includegraphics[width=8cm]{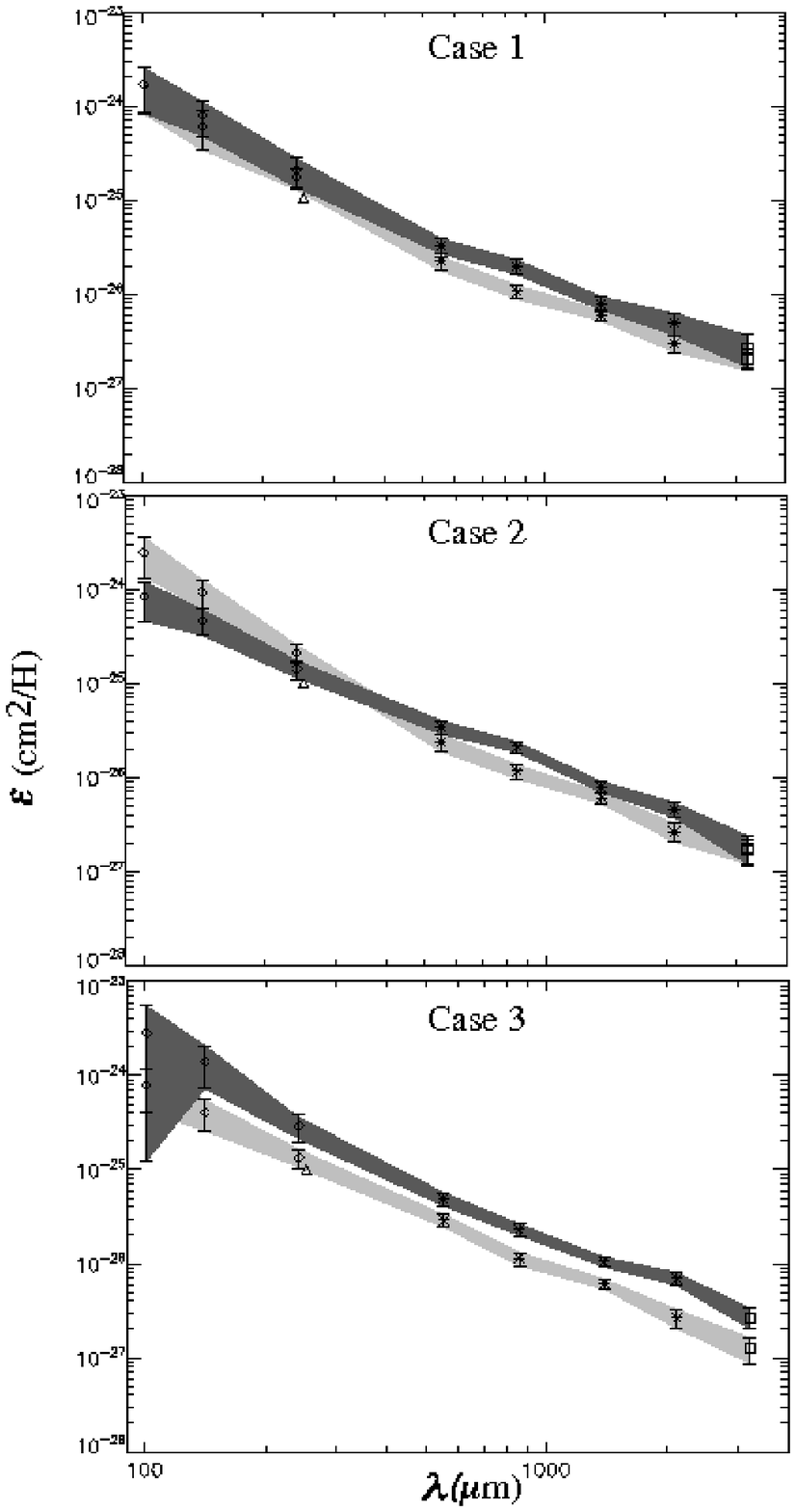}
\caption{Median dust emissivity SEDs for case 1 (\CaseOne: upper panel), case
2 (\CaseTwo: middle panel) and case 3 (\CaseThree: lower panel). The
DIRBE, \Archeops and WMAP data correspond to diamond, star, and square
symbols, respectively. The shaded areas show the $\rm \pm$ 1-$\sigma$
dispersion around each emissivity SED, in dark grey for the atomic
phase and in light grey for the molecular phase.
The triangle symbol
at $\rm \lambda=250\,\mic$ corresponds to the emissivity value derived by
\citet{Boulanger96} for the diffuse atomic medium at 250 $\mic$. \label{fig_spec_med}}
\end{center}
\end{figure}

Dust in dense molecular clouds is expected to be colder than in the
surrounding atomic material, as long as no significant star formation
is occurring in the cloud. However, massive star formation can
significantly heat the dust in molecular clouds. In addition, star
formation may increase significantly the turbulence in the star-forming molecular medium, which in turn, could modify the dust emission
properties. To search for these potential variations, we
classify the regions according to 3 categories based upon the dust
temperature derived for dust in the atomic phase ( $\rm T_d^{HI}$)
and in the molecular phase ($\rm T_d^{CO}$):

\begin{itemize}

\item{\CaseOne: atomic and molecular medium with similar dust
temperatures defined as $\rm T_d^{CO} <T_d^{HI}<T_d^{CO}+0.6\,K$. This
case allows us to compare the dust emissivity in the two environments in
regions where the ISRF intensity is similar in both phases. This
category accounts for 18 regions in our sample.}

\item{\CaseTwo: dust in the molecular phase significantly colder than in the
atomic phase: $\rm T_d^{CO} +2.5\,K <T_d^{HI}<T_d^{CO} + 4\,K$. In this
case, the colder dust temperature in the molecular phase is indicative of a
lower ISRF intensity in the cloud, most likely indicating dense
molecular environment with little star formation. This case is the
most relevant to the search for emissivity variations linked to the
formation of dust aggregates, since so far, these have been found
in such regions \citep[see e.g.,][]{Bernard99,Stepnik03}. This category
accounts for 45 regions in our sample.}

\item{\CaseThree: dust in the atomic phase significantly colder than in the
molecular phase: $\rm T_d^{HI} +2\,K <T_d^{CO}<T_d^{HI} + 4\,K$. This
category should incorporate most of the massive star-forming regions.
The lower temperature (here $\rm T_d^{HI}$) is lower than in the
previous case 2 in order to include enough regions. This category
accounts for 25 regions in our sample.}

Figure\,\ref{fig_regions} shows the location of all studied regions
for the three temperature cases. Temperature histograms are shown in
Fig.\,\ref{fig_hist}, for each case and phase. Although the cases are
selected based on the relative temperature between the two phases, it
can be seen that they also select absolute temperature behaviors,
since the absolute values of the dust temperatures are roughly
reversed in case 2 and 3 ($\rm T_{HI}^{Case2} \simeq T_{CO}^{Case3}
\simeq 17\,K$ and $\rm T_{CO}^{Case2} \simeq T_{HI}^{Case3} \simeq
14\,K$). Appendix (Table 7 and 8) indicates for each case
and studied region, the number of pixels considered, the correlation factors obtained as well
as the dust temperature values, and the emissivity spectral index in both the FIR and the submm range.
\section{Results}
\label{sec:results}

Figure\,\ref{fig_spec_med} shows the SEDs of the median emissivity
derived from the correlation values, for the atomic and molecular
phases, for each of the cases described in Sect.\,\ref{sec:cases}. The
corresponding values are also given in Table \ref{tab_val_med}.

The emissivity values shown in Fig.\,\ref{fig_spec_med} were not
normalised in any other way than assuming the fixed conversion factors
described in Sect.\,\ref{sec:convfact}. In Figure\,\ref{fig_spec_med}, we also show the emissivity value derived from the FIRAS
data by \citet{Boulanger96} for the atomic diffuse ISM in the solar
neighbourhood at $\rm \lambda=250\,\mic$, assuming $\rm \beta=2$. It can be seen
that the emissivity values derived here are in rough agreement with
that value. We note however that the atomic emissivities that we derived
are somewhat higher than in the solar neighbourhood. 
We also note that there is a continuity in the emissivity SED between
the WMAP W band and the \Archeops range indicating that there is no
obvious calibration difference between the two experiments, which are
both calibrated to the CMB dipole in this range.
\end{itemize}

The scaling of emissivity values for the molecular phase can be
modified by assuming a different $\rm X_{CO}$ factor. The $\XCO$ factor is difficult to determine. It is expected
to vary with various factors such as the metallicity and the chemistry
of the ISM, and could therefore change from cloud to
cloud \citep[e.g.,][]{Magnani95}. In Figure\,\ref{fig_spec_med_norm},
we show the same emissivity SEDs, where we have arbitrarily chosen the
$\rm X_{CO}$ value so that the atomic and molecular emissivities match at
the longest wavelengths. For each case, the derived $\rm X_{CO}$ values are
given in the plot. Figure\,\ref{fig_spec_med_norm} more clearly shows
the variations in the spectral shape. From that figure it is apparent 
that in case 1 and 3, the atomic and molecular medium emissivity SEDs
are almost parallel, while in case 2 (colder CO), the molecular
emissivity SED is significantly steeper than the atomic one in the
FIR range. We note however, that if the absolute
level of the emissivities in the molecular phase can vary according to
the assumed $\XCO$ value, the spectral shape of the emissivity SED and therefore the emissivity slope ($\beta$) will remain unchanged.

In Figure\,\ref{fig_spec_med_norm}, we also show power laws with
$\rm \beta=2.5$, $\rm \beta=2$ and $\rm \beta=1.5$ normalised at 100 $\mic$ for the atomic
phase. Although all emissivity SEDs globally show the expected
decrease with wavelength, it can be seen that,
in all 3 cases, the emissivity SED is steeper in the FIR (DIRBE data)
and flattens significantly in the submillimeter above $\rm \lambda >
500\,\mic$ (\Archeops and WMAP data). Tables \ref{tab_beta_med} and
\ref{tab_beta_ratio_med} summarise the median values of the emissivity
spectral index at all wavelengths in the DIRBE and \Archeops
wavelength ranges, respectively. These values were inferred from a
linear adjustment of the emissivity SEDs. Uncertainties correspond to
1-$\rm \sigma$. It can be seen that the emissivity index is generally
higher than $\rm \beta^{fir}\simeq2$ in the FIR corresponding to a steep
emissivity decrease with wavelength, and is about $\rm \beta^{submm}\simeq1.5$
in the submm and mm domain, corresponding to a flatter
spectrum. Apart from this common behavior, there are also significant
differences between the 3 temperature cases, which are discussed
below.

\subsection{Case 1: \CaseOne}
\begin{table*}[!t]
\begin{center}
\begin{tabular}{  p{1.2cm} p{0.6cm}  c  c  c c  c  c }
\hline
\hline
Instrument & $\rm \lambda$ &  \multicolumn{2}{c }{Case 1: \CaseOne} &  \multicolumn{2}{c}{ Case 2: \CaseTwo} &   \multicolumn{2}{c }{Case 3: \CaseThree} \\
& ($\rm \mic$)& $\rm \epsilon_{HI}$ & $\rm \epsilon_{CO}$ & $\rm \epsilon_{HI}$ & $\rm \epsilon_{CO}$ & $\rm \epsilon_{HI}$ & $\rm \epsilon_{CO}$ \\
\hline
DIRBE & 100 & 1.7 $\pm$ 0.9 10$^{-24}$ & 1.7 $\pm$ 0.9 10$^{-24}$ & 8.4 $\pm$ 3.8 10$^{-25}$ & 2.4 $\pm$ 1.1 10$^{-24}$ & 2.8 $\pm$ 2.7 10$^{-24}$ & 7.8 $\pm$ 3.8 10$^{-25}$\\
DIRBE & 140 & 8.0 $\pm$ 3.2 10$^{-25}$ & 6.2 $\pm$ 2.8 10$^{-25}$ & 4.7 $\pm$ 1.5 10$^{-25}$ & 9.4 $\pm$ 3.3 10$^{-25}$ & 1.4 $\pm$ 0.7 10$^{-24}$ & 4.0 $\pm$ 1.5 10$^{-25}$\\
DIRBE & 240 & 2.1 $\pm$ 0.7 10$^{-25}$ & 1.8 $\pm$ 0.5 10$^{-25}$ & 1.4 $\pm$ 0.3 10$^{-25}$ & 2.1 $\pm$ 0.5 10$^{-25}$ & 2.9 $\pm$ 0.9 10$^{-25}$ & 1.3 $\pm$ 0.3 10$^{-25}$ \\
\Archeops & 550 & 3.3 $\pm$ 0.7 10$^{-26}$ & 2.2 $\pm$ 0.5 10$^{-26}$ & 3.4 $\pm$ 0.6 10$^{-26}$ & 2.4 $\pm$ 0.5 10$^{-26}$ & 4.9 $\pm$ 0.8 10$^{-26}$ & 2.9 $\pm$ 0.5 10$^{-26}$\\
\Archeops & 850 & 2.0 $\pm$ 0.4  10$^{-26}$ & 1.1 $\pm$ 0.2  10$^{-26}$ & 2.1 $\pm$ 0.3  10$^{-26}$ &  1.2 $\pm$ 0.2  10$^{-26}$ & 2.3 $\pm$ 0.3 10$^{-26}$  & 1.1 $\pm$ 0.2 10$^{-26}$ \\
\Archeops & 1400 & 8.0 $\pm$ 1.5 10$^{-27}$ & 6.0 $\pm$ 0.8 10$^{-27}$ & 7.9 $\pm$ 1.1 10$^{-27}$ & 6.1 $\pm$ 0.9 10$^{-27}$ & 1.0 $\pm$ 0.1 10$^{-26}$ & 6.1 $\pm$ 0.8 10$^{-27}$\\
\Archeops & 2100 & 5.0 $\pm$ 1.4 10$^{-27}$ & 3.1 $\pm$ 0.6 10$^{-27}$ & 4.6 $\pm$ 0.9 10$^{-27}$ & 2.6 $\pm$ 0.6 10$^{-27}$ & 6.9 $\pm$ 1.1 10$^{-27}$ & 2.6 $\pm$ 0.6 10$^{-27}$ \\
WMAP & 3200 & 2.7 $\pm$ 1.0 10$^{-27}$ & 2.1 $\pm$ 0.5 10$^{-27}$ & 1.8 $\pm$ 0.6 10$^{-27}$ & 1.7 $\pm$ 0.5 10$^{-27}$ & 2.7 $\pm$ 0.7 10$^{-27}$ & 1.2 $\pm$ 0.4 10$^{-27}$ \\
\hline
\end{tabular}
\end{center}
\caption{Median values of the dust emissivity in the molecular and the
atomic phase for each studied case: case 1 (\CaseOne), case 2
(\CaseTwo), case 3 (\CaseThree).  Error bars represent the $\pm$
1-$\rm \sigma$ uncertainty estimates.\label{tab_val_med} }
\end{table*}

\begin{table*}[!t]
\begin{center}
\begin{tabular}{ c  c  c  c c  c  c }
\hline
\hline
$\rm \lambda$ &  \multicolumn{2}{c }{Case 1: \CaseOne} &  \multicolumn{2}{c }{ Case 2: \CaseTwo} &   \multicolumn{2}{c }{Case 3: \CaseThree} \\
 ($\rm \mic$)& $\rm \beta_{HI}$ & $\rm \beta_{CO}$ & $\rm \beta_{HI}$ & $\rm \beta_{CO}$ & $\rm \beta_{HI}$ & $\rm \beta_{CO}$ \\
\hline
 100 & 2.3 $\pm$ 1.7 & 3.0 $\pm$ 2.3 & 1.7 $\pm$ 1.9 & 2.9 $\pm$ 2.1 & 2.1 $\pm$ 3.3 & 2.0 $\pm$ 1.7\\
 140 & 2.4 $\pm$ 0.5 & 2.5 $\pm$ 0.6 & 2.1 $\pm$ 0.6 & 2.8 $\pm$ 0.6 & 2.8 $\pm$ 0.9 & 2.1 $\pm$ 0.5 \\
 240 & 2.3 $\pm$ 0.3 & 2.5 $\pm$ 0.3 & 1.8 $\pm$ 0.2  & 2.7 $\pm$ 0.3 & 2.3 $\pm$ 0.3 & 1.9 $\pm$ 0.2 \\
 550 & 1.9 $\pm$ 0.3 & 2.1 $\pm$ 0.2 & 1.5 $\pm$ 0.2 & 2.3 $\pm$ 0.2 & 2.0 $\pm$ 0.3 & 2.0 $\pm$ 0.3 \\
 850 & 1.6 $\pm$ 0.3 & 1.3 $\pm$ 0.2 & 1.5 $\pm$ 0.2 & 1.5 $\pm$ 0.2 & 1.7 $\pm$ 0.2  & 1.6 $\pm$ 0.2  \\
 1400 & 1.5 $\pm$ 0.4 &  1.3 $\pm$ 0.2 & 1.6 $\pm$ 0.3 & 1.5 $\pm$ 0.3 & 1.3 $\pm$ 0.2 & 1.5 $\pm$ 0.3 \\
 2100 & 1.2 $\pm$ 0.2 & 1.3 $\pm$ 0.2 & 1.7 $\pm$ 0.3 & 1.6 $\pm$ 0.3 & 1.3 $\pm$ 0.3 & 1.7 $\pm$ 0.3  \\
 3200 & 1.2 $\pm$ 0.6 & 0.8 $\pm$ 0.5 & 2.1 $\pm$ 0.7  & 1.3 $\pm$ 0.8 & 2.2 $\pm$ 0.6  & 1.4 $\pm$ 0.8 \\
\hline
\end{tabular}
\end{center}
\caption{Median values of the emissivity spectral index ($\rm \beta$) in the
molecular and the atomic phase for each studied case and each
wavelength.  Error bars represent the 1-$\rm \sigma$ uncertainty
estimates.\label{tab_beta_med} }
\end{table*}
\begin{table*}[!t]
\begin{center}
\begin{tabular}{  c  c  c  c  c  }
\hline
\hline
$\rm \lambda$ Range  & Phase & Case 1: \CaseOne & Case 2: \CaseTwo&  Case 3: \CaseThree \\
\hline
$\rm \beta$ [100-240] $\rm \mic$ & HI & 2.4 $\pm$ 0.5 & 2.1 $\pm$ 0.6 & 2.8 $\pm$ 0.9 \\
  & CO & 2.5 $\pm$ 0.6 & 2.8 $\pm$ 0.6 & 2.1 $\pm$ 0.5 \\
$\rm \beta_{CO}/\beta_{HI}$ && 1.0  $\pm$ 0.5 & 1.3 $\pm$ 0.5 & 0.7 $\pm$ 0.6 \\

$\beta$ [550-2100] $\mic$ & HI &  1.5 $\pm$ 0.2 & 1.5 $\pm$ 0.2 & 1.5 $\pm$ 0.1\\
  & CO & 1.4 $\pm$ 0.2 & 1.5 $\pm$ 0.2 & 1.7 $\pm$ 0.2\\
$\rm \beta_{CO}/\beta_{HI}$& & 0.9  $\pm$ 0.3 & 1.0 $\pm$ 0.2 & 1.1 $\pm$ 0.2 \\
\hline
\end{tabular}
\caption{Median values of the emissivity spectral index ($\beta$) for the
FIR and submm domain, for each studied cases. Errors bars represent
the 1-$\rm \sigma$ uncertainty estimates. \label{tab_beta_ratio_med}}
\end{center}
\end{table*}
In this case, the emissivity SED of the dust associated with the atomic
and molecular phases are almost parallel, and both clearly experience
the same slope change between the DIRBE and the \Archeops
wavelengths. This similarity probably indicates that, in this case,
the dust emission properties associated with the molecular and atomic
phases are the same.

In addition, the absolute emissivity values in the two phases are
relatively close. This indicates that our adopted $\XCO$ value is
reasonable in that case. There is no reason in principle that the
dust abundance $\rm X_{d}$ could differ between the two
phases. Therefore, an incorrect estimate of the $\XCO$ parameter would produce
a wavelength-independent difference between the two emissivity
SEDs. Looking at the median values of
the emissivity ratios in the two phases presented in
Table\,\ref{tab_em_ratio_med}, we can see that the emissivity in the
atomic phase is higher that in the molecular phases at all
wavelengths, by about 30$\%$ for the assumed $\XCO$. This difference
could be accounted for by a change in the $\XCO$ value, which would
then have to be slightly lower than the one assumed, $\rm \XCO=1.6\times 10^{20}\XCOUNIT$.

\subsection{Case 2: \CaseTwo}
\label{sec_case2}

This case includes regions where dust in the molecular
phase is significantly colder than in the surrounding atomic
medium. It is of the greatest interest when comparing our
results with those obtained using the PRONAOS data in the Taurus
and the Polaris regions, which lead previous authors to propose the
presence of fractal dust aggregates in the molecular phase. Our
results (see Fig.\,\ref{fig_spec_med}) indicate that
the emissivity SED of dust associated with the atomic and molecular medium differ. In the FIR (DIRBE
wavelength range), the emissivity spectral index in the molecular phase
is significantly higher than in the atomic phase. According to
Table\,\ref{tab_beta_ratio_med} the molecular to atomic emissivity
ratio ($\rm \frac{\beta_{CO}}{\beta_{HI}}$) is equal to 1.3 in the
FIR domain.
However, in the submillimeter (\Archeops wavelength range), the
emissivity spectra in the two phases become parallel again, and have quite comparable emissivity values for the assumed value of $\XCO$. Matching exactly the submm emissivities in the two phases in the
submillimeter range would lead to $\rm \XCO=1.5\times10^{20}\XCOUNIT$.
According to Table\,\ref{tab_beta_ratio_med}, $\rm
\frac{\beta_{CO}}{\beta_{HI}}$=1.0, indicating that the
slope of the submm emissivity is not very different from case 1, whose ratio is 0.9.
For the adopted $\XCO$, we observe that the emissivity in the molecular
phase is higher than that in the atomic phase in the FIR range, by
a factor ranging from 3.1 at 100 $\rm \mic$, to about 1.5 at 240 $\rm
\mic$ (see Table \ref{tab_em_ratio_med}). The absolute emissivity
values for the two phases tend to become progressively 
similar around 550 $\mic$. This behavior cannot be attributed to the
assumed $\XCO$ value, which would affect emissivity values at all
wavelengths. We also considered the possibility that this could be due
to an error on the determination of the dust temperature in the
molecular phase. Obtaining the same emissivity slope in the FIR in the
two phases would require an increase of the temperature in the
molecular phase of 2-3 K, which is larger than the temperature
uncertainty, which are between 0.25 K and 0.72 K. In our analysis, we
have so far used the major simplification of a single dust temperature
along the line of sight. In Sect.\,\ref{sec_bias}, we investigate the
effect of this hypothesis on the results. In particular, we consider
the effect of the temperature mixing due to the grain size
distribution, the LOS mixing of the radiation field and the grain
composition.

In this case, the dust temperature in the molecular phase is higher than the
dust temperature in the atomic phase. Taking into account the error bars, the behavior of the
emissivity spectrum is qualitatively the same as in case 1. 
For the assumed $\XCO$, we note that the atomic emissivities are significantly higher
than the molecular ones at all wavelengths. To reproduce the
emissivity in the submm, we would require Xco=$\rm
1.1\times10^{20}\XCOUNIT$. The results of this case show that when the
dust temperature in the atomic phase is close to or colder than the
molecular phase, grains emission properties seem to be the same. 

\section{Possible biases caused by assumptions}
\label{sec_bias}

In the study presented above, we assumed a single dust temperature
along each LOS. This is probably overly simplistic, since we
expect a mixture of temperatures for various reasons. First, dust exhibits a size distribution, and dust temperatures depend
upon grain size for a given radiation field. Second, the ISRF strength probably varies along any given LOS within the Galactic plane, even
towards the external Galactic regions studied here. Third, composition
variations in the dust could produce temperature variations, which
could also affect our results. Since we divide the observed sky
brightness by the black-body function at a single temperature, slight
systematic biases in the dust temperature caused by these effects
could in principle explain the inferred variations of the dust
emissivity, particularly in the FIR region, where the black-body
function is non-linear with respect to the dust temperature.
\begin{table}[!t]
\begin{center}
\begin{tabular}{  c  c  c  c }
\hline
\hline
$\rm \lambda$ ($\mic$)& $\rm \epsilon_{CO}/\epsilon_{HI}$ & $\rm \epsilon_{CO}/\epsilon_{HI}$  &  $\rm \epsilon_{CO}/\epsilon_{HI}$\\
 & Case 1: \CaseOne & Case 2: \CaseTwo & Case 3: \CaseThree \\
\hline
 100 & 0.8 $\pm$ 0.7  & 3.1 $\pm$ 2.9   & 0.2 $\pm$ 0.2 \\
 140 & 0.7 $\pm$ 0.5  & 2.1 $\pm$ 1.6 & 0.2 $\pm$ 0.2 \\
 240 & 0.8 $\pm$ 0.4  & 1.5 $\pm$ 0.8 & 0.4 $\pm$ 0.2  \\
 550 & 0.8 $\pm$ 0.3  & 0.7 $\pm$ 0.2  & 0.6 $\pm$ 0.2 \\
 850 & 0.6 $\pm$ 0.2  & 0.5 $\pm$ 0.2  & 0.5 $\pm$ 0.2 \\
 1400 & 0.8 $\pm$ 0.2 & 0.8 $\pm$ 0.2  & 0.5 $\pm$ 0.1  \\
 2100 & 0.7 $\pm$ 0.2 & 0.6 $\pm$ 0.3  & 0.3 $\pm$ 0.2  \\
 3200 & 0.7 $\pm$ 0.2 & 0.7 $\pm$ 0.4 & 0.3 $\pm$ 0.3 \\
\hline
\end{tabular}
\end{center}
\caption{Median values of the ratio between the dust emissivity in the
molecular phase and in the atomic phase for each studied case.  Error
bars represent the $\pm$ 1-$\sigma$ dispersion around each
value.\label{tab_em_ratio_med}
}
\end{table}
\begin{figure}[!t]
\begin{center}
\includegraphics[width=8cm]{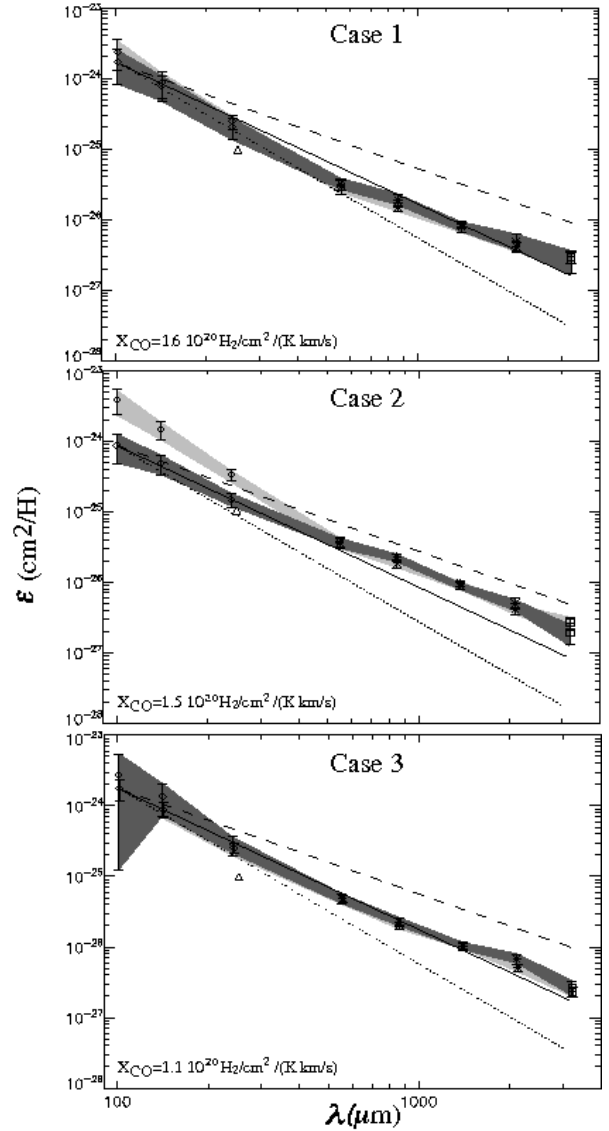}
\caption
{Same as Fig.\,\ref{fig_spec_med} but with the molecular emissivity
scaled to match that of the atomic phase in the range
$\rm 550\,\mic<\lambda<3\,mm$. The corresponding $\rm X_{CO}$ values are given
in the figure. For comparison, power law spectra with $\rm \beta=1.5$
(dashed), $\rm \beta=2.0$ (solid) and $\rm \beta=2.5$ (dot) are shown,
normalised to the atomic emissivity at 100 $\mic$.\label{fig_spec_med_norm}}
\end{center}
\end{figure}

In this section, we explore those 3 possibilities and test the
robustness of our results to temperature variations along the LOS induced by either grain size distribution, ISRF
strength mixing or grain composition variations. In all cases, we produce
predicted emission SEDs using a pertinent dust emission model, which
includes the additional complexity, apply the same treatment applied to the
true data to the modelled SEDs, and compare the derived emissivity
curves with that of the model.
\subsection{Case 3: \CaseThree}

\begin{table*}
\begin{center}
\begin{tabular}{   c  c c  c  c  c  c  c  c  c  }
\hline
\hline
 $\rm \lambda$ ($\rm \mic$) & \multicolumn{3}{c }{Grain size distribution}& \multicolumn{3}{c }{ISRF strength mixture} &  \multicolumn{3}{c }{Grain composition}  \\
 & X$\rm _{ISRF}$=0.05 & X$\rm _{ISRF}$= 0.5& X$\rm _{ISRF}$=4& $\rm \alpha$=2.5 &$\rm \alpha$= 2& $\rm \alpha$= 1.25& T$\rm _2$=10 K & T$\rm _2$= 15 K & T$\rm _2$=22 K \\
\hline
 100 & 1.13 & 0.89 & 0.93 & 1.20 & 1.46 & 3.06 & 1.06 & 1.29 & 1.16 \\
 140  & 1.13 & 0.96 & 0.98 & 1.23 & 1.45 & 2.66 & 1.12 & 1.27 & 1.14 \\
 240  & 1.11 & 1.00 & 1.02 & 1.17 & 1.29 & 1.84 & 1.03 & 1.19 & 1.13 \\
 550  & 1.07 & 1.03 & 1.03 & 1.09 & 1.12 & 1.27 & 0.96 & 1.05 & 1.04 \\
 850  & 1.04 & 1.02 & 1.02 & 1.05 & 1.07 & 1.14 & 1.07 & 1.10 & 1.08 \\
 1400  & 1.00 & 0.99 & 0.99 & 1.01 & 1.01 & 1.05 & 1.00 & 1.04 & 1.05 \\
 2100  & 0.99 & 0.99 & 0.99 & 0.99 & 1.00 & 1.01 & 0.98 & 1.01 & 1.03 \\
\hline
\end{tabular}
\caption{Ratios of the predicted to apparent emissivities, using
the same processing as for the data with different models. Cols 2-4: using the \citet{Desert90} model at various ISRF strength ($\rm X_{ISRF}$). Cols 5-7:
assuming a mixture of the radiation field intensity on each LOS, using
the model by \citet{Dale01} for various values of
the mixing parameter $\rm \alpha$. Cols 8-10: using the two component model of
\citet{Finkbeiner99} for different temperatures of the warm component
(T$\rm _2$).\label{tab_effects}}
\end{center}
\end{table*}

\subsection{Effect of the grain size distribution}
\label{sec:grain_size}
\begin{figure*}[!t]
\begin{center}
\includegraphics[width=18cm]{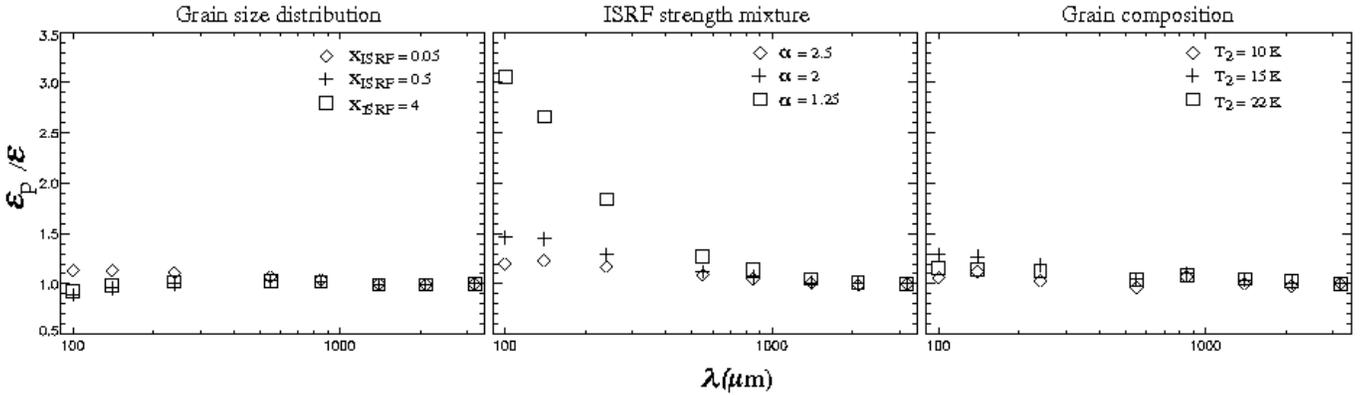}
\caption{\label{fig_em_models}
Ratios of the predicted ($\rm \epsilon_p$) to apparent
emissivities ($\rm \epsilon$), using the same processing as for the
data with different modelled SEDs. Left panel: using the
\citet{Desert90} model at various ISRF strengths ($\rm
X_{ISRF}$). Middle panel: assuming a mixture of the radiation field
intensity on each line of sight, using the model description proposed by
\citet{Dale01} for various values of the mixing parameter $\rm
\alpha$. Right panel: using the two component model of
\citet{Finkbeiner99} for different temperatures of the warm component
(T$\rm _2$).}
\end{center}
\end{figure*}

To test the influence of temperature caused by the BG
size distribution, we generated model SEDs at various ISRF intensities
($\rm X_{ISRF}$), using the dust model of \citet{Desert90}, which
includes a realistic size distribution for large grains and computes
the induced temperature mixing, by taking into account the effects of
grain temperature fluctuations. Model SEDs were computed in all
instrument filters used in our study, using the proper color
correction to match the flux convention of each instrument. We then
applied the same processing to the model SEDs as applied to the actual
data. As the model assumes a dust emissivity with $\rm \beta=2$ over
the entire wavelength range considered here, we expect in that case to recover a
simple power law with slope $\rm \beta=2$. We performed
the test over $\rm X_{ISRF}$ values ranging from 0.05 to 10 times the
solar neighbourhood ISRF strength. Ratios of the predicted
emissivity values inferred from the model and the apparent emissivity
obtained with our procedure are given in Table\,\ref{tab_effects},
normalised to the emissivity value in the WMAP band. Figure
\ref{fig_em_models} is a graphical representation of
Table \ref{tab_effects}. The left panel shows that the emissivity
ratio is close to 1. We therefore recover emissivity spectra very
similar to the input power law, indicating that no significant bias is
introduced by not considering the LOS temperature mixing induced by a
realistic dust size distribution. We also checked the influence of
changing the spectral shape of the ISRF, in particular increasing the
UV content as expected close to young stars. This test showed that the
shape of the ISRF does not affect our results, because BGs absorb uniformly across the entire visible to
ultra-violet range of the model used. We conclude that the method
described above introduces no significant bias in that case.

\subsection{Effect of the ISRF strength mixture}
\label{sec:isrf}
\begin{figure}
\begin{center}
\includegraphics[width=8cm]{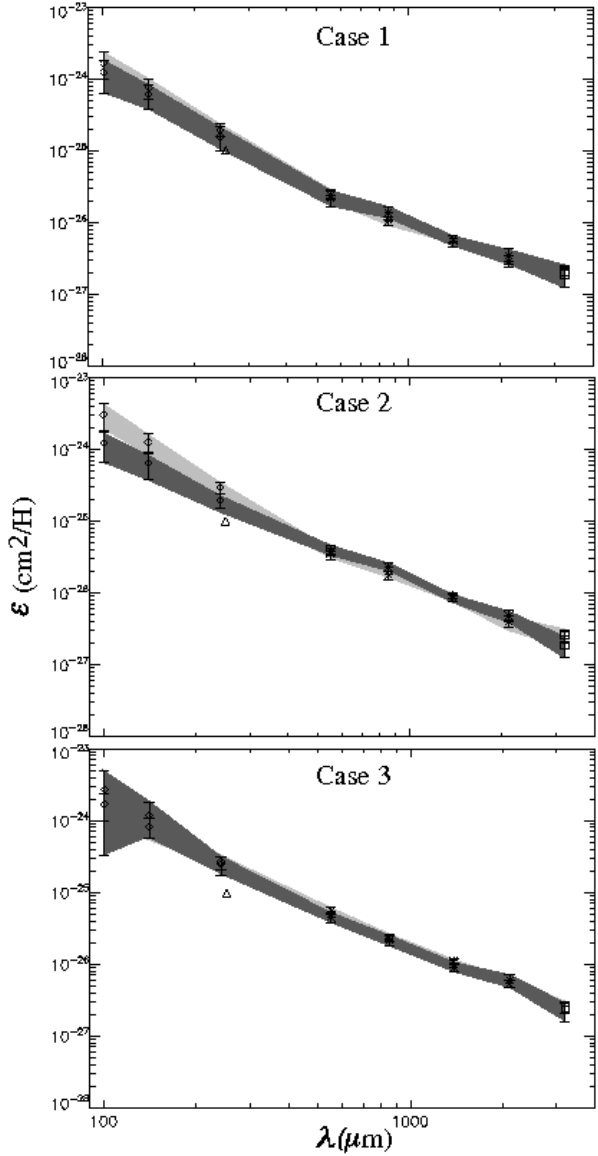}
\caption{\label{fig_spec_med_corr} Median dust emissivity SEDs for the 3
cases (see Fig.\,\ref{fig_spec_med} for the description of the curves)
corrected for the hypothesis of a mixture of different interstellar
radiation field along the line of sight. The molecular emissivity
has been scaled to match that of the atomic phase in the range
$\rm 550\,\mic<\lambda<3\,mm$. The triangle corresponds to
the emissivity value derived by \citet{Boulanger96} in the diffuse
medium.}
\end{center}
\end{figure}

To test the influence of the variations in the ISRF strength
mixture along the LOS, we use the model proposed by \citet{Dale01},
who introduced the concept of local SED combination, assuming a
power-law distribution of dust mass subjected to a given heating
intensity $\rm dM_d(X_{ISRF})$, to interpret the emission by
external galaxies:

\begin{equation}
dM_d(X_{ISRF}) \propto X_{ISRF}^{-\alpha}dX_{ISRF},
\end{equation}
where $\rm \alpha$ controls the relative contribution of the various
ISRF intensities ($\rm X_{ISRF}$) to the SEDs. \citet{Dale01} used
ISRF strengths values in the range $\rm 0.3<X_{ISRF}<10^5$ and derived
$\rm \alpha$ values in the range $\rm 1<\alpha<2.5$ for external
galaxies. We first computed emission spectra using the
\citet{Desert90} model at various $\rm X_{ISRF}$ values (noted
$\rm I^{mod}_{\nu}(X_{ISRF})$), then summed them over the same $\rm
X_{ISRF}$ range as proposed by \citet{Dale01}, according to
\begin{equation}
I^{tot}_{\nu}=\frac{\sum_{X_{ISRF}} I^{mod}_{\nu}(X_{ISRF}) \times X_{ISRF}^{-\alpha}}{\sum_{X_{ISRF}} X_{ISRF}^{-\alpha}}.
\end{equation}
In Figure\,\ref{fig_em_models} (middle plot), it can be seen that
significant departures from model predictions can occur in the FIR, close to the peak of the
emission, for flat $\rm X_{ISRF}$ distributions (i.e., low $\rm \alpha$
values). These deviations become progressively smaller towards steeper
distribution, as expected. We note that the apparent dust temperature
derived from the fit decreases as $\rm \alpha$ increases, which can be
used to constrain $\rm \alpha$ values. From the model predictions, we
corrected our emissivity estimates, using the apparent dust temperature
for each measured SED to infer the true $\rm \alpha$ value and then deduce the factor we have to apply to correct our values, knowing the ratio of the predicted to apparent emissivity. In general, assuming a single dust temperature along the LOS causes the emissivity values between 100 and 550 $\rm \mic$ to be underestimated. 
For most of the regions considered here, the apparent temperature is
moderate, which is indicative of steep $\rm X_{ISRF}$ mixing with $\rm \alpha$
close to 2.5.
\begin{table*}[!t]
\begin{center}
\begin{tabular}{  c c  c  c c  c  c }
\hline
\hline
 $\rm \lambda$ &  \multicolumn{2}{c }{Case 1: \CaseOne} &  \multicolumn{2}{c }{ Case 2: \CaseTwo} &   \multicolumn{2}{c }{Case 3: \CaseThree} \\
 ($\rm \mic$)& $\rm \epsilon_{HI}$ & $\rm \epsilon_{CO}$ & $\rm \epsilon_{HI}$ & $\rm \epsilon_{CO}$ & $\rm \epsilon_{HI}$ & $\rm \epsilon_{CO}$ \\
\hline
 100 & 1.8 $\pm$ 0.9 10$^{-24}$ & 1.6 $\pm$ 0.7 10$^{-24}$ & 1.2 $\pm$ 0.6 10$^{-24}$ & 2.0 $\pm$ 0.9 10$^{-24}$ & 2.7 $\pm$ 2.5 10$^{-24}$ & 8.5 $\pm$ 4.5 10$^{-25}$\\
 140 & 8.7 $\pm$ 3.3 10$^{-25}$ & 7.5 $\pm$ 2.6 10$^{-25}$ & 6.1 $\pm$ 2.6 10$^{-25}$ & 8.2 $\pm$ 2.8 10$^{-25}$ & 1.3 $\pm$ 0.7 10$^{-24}$ & 4.6 $\pm$ 1.8 10$^{-25}$\\
 240 & 2.3 $\pm$ 0.8 10$^{-25}$ & 1.9 $\pm$ 0.4 10$^{-25}$ & 1.8 $\pm$ 0.5 10$^{-25}$ & 1.9 $\pm$ 0.4 10$^{-25}$ & 2.8 $\pm$ 0.8 10$^{-25}$ & 1.5 $\pm$ 0.3 10$^{-25}$ \\
 550 & 3.3 $\pm$ 0.8 10$^{-26}$ & 2.4 $\pm$ 0.5 10$^{-26}$ & 3.9 $\pm$ 0.6 10$^{-26}$ & 2.3 $\pm$ 0.5 10$^{-26}$ & 5.1 $\pm$ 0.8 10$^{-26}$ & 3.0 $\pm$ 0.5 10$^{-26}$\\
 850 & 2.0 $\pm$ 0.4  10$^{-26}$ & 1.1 $\pm$ 0.1  10$^{-26}$ & 2.3 $\pm$ 0.3  10$^{-26}$ &  1.2 $\pm$ 0.2  10$^{-26}$ & 2.4 $\pm$ 0.3 10$^{-26}$  & 1.2 $\pm$ 0.2 10$^{-26}$ \\
 1400 & 7.9 $\pm$ 1.5 10$^{-27}$ & 5.9 $\pm$ 0.8 10$^{-27}$ & 8.1 $\pm$ 1.1 10$^{-27}$ & 5.9 $\pm$ 0.8 10$^{-27}$ & 1.0 $\pm$ 0.1 10$^{-26}$ & 6.0 $\pm$ 0.8 10$^{-27}$\\
 2100 & 5.0 $\pm$ 1.4 10$^{-27}$ & 3.0 $\pm$ 0.6 10$^{-27}$ & 4.6 $\pm$ 0.9 10$^{-27}$ & 2.6 $\pm$ 0.6 10$^{-27}$ & 6.8 $\pm$ 1.1 10$^{-27}$ & 2.6 $\pm$ 0.6 10$^{-27}$ \\
 3200 & 2.7 $\pm$ 1.0 10$^{-27}$ & 2.0 $\pm$ 0.5 10$^{-27}$ & 1.8 $\pm$ 0.6 10$^{-27}$ & 1.7 $\pm$ 0.5 10$^{-27}$ & 2.7 $\pm$ 0.7 10$^{-27}$ & 1.2 $\pm$ 0.4 10$^{-27}$ \\
\hline
\end{tabular}
\end{center}

\caption{Median values of the dust emissivity in the molecular and the
atomic phase for each studied case, after correction of a mixture of
the radiation field intensity along the line of sight. Error bars
represent the $\pm$ 1-$\sigma$ uncertainty estimates.\label{tab_val_med_corr}
}
\end{table*}
The resulting corrected median emissivity values are shown in
Fig.\,\ref{fig_spec_med_corr}, after renormalisation by $\rm X_{CO}$. Comparison with
Fig.\,\ref{fig_spec_med_norm} shows that $\rm X_{ISRF}$ mixing has little
impact on our results. In particular, both the break in the emissivity
slope near 500 $\mic$ and the dust emissivity increase in
the molecular phase in case 2 are still present. We note that the amplitude of
the emissivity increase is slightly lower after the correction, as
shown in Table \ref{tab_val_med_corr}, with a median ratio of 2.0
$\pm$ 1.6 between the 2 spectra at 100 $\mic$ after correction,
instead of 3.1, before renormalisation by $\rm X_{CO}$. The emissivity slopes in the FIR are equal to 2.7
$\pm$ 0.6 and 2.3 $\pm$ 0.6, in the FIR for the molecular and atomic
phase, respectively, and 1.5 $\pm$ 0.2 and 1.6 $\pm$ 0.2 in the
submm.

\subsection{Effect of the grain composition}
\label{sec:composition}

We use the \citet{Finkbeiner99} model to test the influence of dust
composition on our results. In the framework of this model, the shape
of the submm emission is assumed to be caused by a mixture of
silicate and graphite grains that reach different temperatures
(T$\rm _1$ and T$\rm _2$ for the silicate and graphite components,
respectively), the temperatures being linked by the
radiation field. In principle, this could also affect our
interpretation, since we assumed a single dust temperature to derive
the dust emissivity spectrum. Unlike \citet{Finkbeiner99} however, we
used the same $\rm \lambda^{-2}$ power law emissivity for each grain
component, to simplify the comparison between the recovered
emissivity and the model. We computed the emissivity ratio at various
temperatures of the graphitic component ranging from 8 K to 22 K (see
the right plot of Fig.\,\ref{fig_em_models}). Table\,\ref{tab_effects}
shows that in this case the bias introduced by our simplification
hypothesis does not significantly impact our results. Similar tests using various emissivity power laws for the two components gave equally good results.

\section{Discussion}
\label{sec:discussion}
Regarding the absolute values of the derived emissivities, it can be
seen in Fig.\,\ref{fig_spec_med} that the values we derived for the
atomic phase are in rough agreement with that derived by
\citet{Boulanger96} ($\rm 10^{-25}$ cm$\rm ^2$/H at 250 $\rm
\mic$). This agreement indicates that the regions we have selected in
this study, which are mainly located towards the Galactic $\rm 2^{nd}$
quadrant, have similar dust content and properties to the
solar neighbourhood. A closer inspection of Fig.\,\ref{fig_spec_med}
shows that our emissivity values in the atomic phase are
systematically higher than the solar neighbourhood values by a factor
of 1.9, 1.3, and 2.6 for cases 1, 2, and 3, respectively. The corresponding
median dust temperatures derived are 15.0 K, 16.9 K, and 14.0 K
respectively (see Table 7), which are also lower than the temperature
inferred by \citet{Boulanger96} for the diffuse ISM in the solar
neighbourhood (17.5 K).

We also emphasize that the atomic and molecular emissivity spectra for
cases 1 and 3 could be reproduced over the full wavelength range
studied by slightly decreasing the $\XCO$ value used for the column
density conversion. We therefore consider it likely that the actual
$\XCO$ value is slightly lower than the assumed value of $\rm
2.3\times 10^{20}\,H_2/cm^2/(K\,km/s)$ in those cases.
In case 2, however, the emissivity spectra have a
significantly different shape and cannot be reproduced over the full
wavelength range by simply changing the $\XCO$ value. We note however that a
similar decrease to that in case 1 would improve the match in the
submm and millimeter range.

Regarding the spectral shape of the emissivity, the results shown in
Sect.\,\ref{sec_case2} indicate that the FIR dust emissivity of dust
associated with molecular regions is significantly higher than that of
the surrounding atomic medium, in cases where the dust temperature is colder in the molecular phase than in the atomic phase (Case 2). Taking into account the error bars derived in Sect.\,\ref{sec:method}
and shown in Fig\,\ref{fig_spec_med_norm},
this result is significant at the $\rm 5-\sigma$ level.
This result is in qualitative
agreement with the result obtained by \citet{Stepnik03}, who showed
that an increase in the dust emissivity by a factor of 3-4 in the molecular
phase is required to explain the low dust temperature in that phase.
We note that their study was limited to
the wavelength range from 200 to 600 $\rm \mic$.
The increase that we found for the DIRBE wavelength range is
therefore of the same order, although slightly smaller, than that
derived by \citet{Stepnik03} for a particular cold molecular filament
in Taurus. In our case 2 regions, we note that on average the dust is
3.2 K colder in the molecular phase than in the atomic phase. This is
also similar to, although slightly lower than, the difference observed
by \citet{Stepnik03} in the Taurus filament (4.8 K). Our result
therefore confirms the hypothesis proposed by \citet{Bernard99} and
\citet{Stepnik03} that dust aggregation may be responsible for the
decrease in the dust equilibrium temperature. In quiescent molecular clouds of case 2, aggregation of BGs in
fractal clusters is a likely cause of the FIR emissivity increase, the
corresponding decrease in the dust equilibrium temperature being a
consequence of the higher emissivity and more efficient cooling.

\citet{Stepnik03} inferred an increase in emissivity by
a constant factor with wavelength. Their data did not allow them to
measure directly the spectral shape of the emissivity, and they used
an average $\rm \beta$ index over the FIR wavelength range of the PRONAOS
experiment. Their approximation of a wavelength independent increase
in the emissivity was supported by the results of discrete dipole
approximation calculations for dust aggregates \citep[e.g., ][]{Bazell90}
that predict a constant absorptivity increase for fractal aggregates
all the way into the millimeter. Our results however show that the
emissivity spectrum in the molecular phase follows the same power law as in
the atomic phase. Although the dust emissivity in the cold molecular
phase is significantly higher in the FIR, it recovers the atomic
phase values in the submm and mm ranges.

We note however that reproducing the temperature difference by
increasing the dust emissivity only requires the emissivity to be higher in
the FIR, where large grains emit most of their energy. Therefore, the close agreement between the emissivities derived for the submm and mm does not therefore the thermal argument upon which the conclusion of dust aggregates is based. A simple calculation of the thermal
equilibrium shows that the measured emissivity increase
in case 2 would produce a temperature decrease of about 2 K, assuming that
the increase below $\rm \lambda = 100\,\mic$ equals that at 100 $\mic$. This
is slightly lower than the observed temperature decrease, which could
indicate that the emissivity continues increasing below 100 $\mic$.

In all cases studied, the emissivity spectra derived are steeper in
the FIR, with spectral index of $\rm \beta \simeq 2.4$ and flatten in
the submm and mm regime, where the spectral index reaches $\rm \beta
\simeq 1.5$. Taking into account, the error bars derived in
Sect.\,\ref{sec:method} and shown in Fig.\,\ref{fig_spec_med_norm},
this result is significant at the $\rm 10-\sigma$ and $\rm 15-\sigma$ level
for cases 1 and 2, respectively. We note that attributing the flattening
to calibration errors would require the calibration of both the
Archeops and WMAP experiments to be wrong by more than a factor of 2
in the same direction, which is very unlikely. This behavior could
in principle be due to the presence of two dust components at
different equilibrium temperatures, as proposed in
phenomenological models with two temperature components
\citep[e.g.,][]{Finkbeiner99}. Such models have been invoked to explain
the flatness of submm emission spectrum, but require the existence of
a cold component with $\rm T_{d}\simeq\,9\,K$, whose origin in the
diffuse ISM remains to be understood. However, we note that the
spectral indices that we obtained in this work are in good agreement with those
proposed in the \citet{Finkbeiner99} model.

A more physical approach consists of interpreting the submm flattening
by intrinsic dust properties. To explain the FIR/mm dust emission, \citet{Meny07} proposed a model based on some specific properties of the
amorphous state. First, they considered a temperature-independent
emission caused by excitation of acoustic lattice vibrations, coming from
the coupling between the electromagnetic fields and a disordered
charge distribution (DCD) that characterizes the amorphous nature of
dust. Their description also takes into account resonant absorption
and relaxation processes associated with localized asymmetric two-level
systems (TLS) in the grains, which produce additional emission
with emissivity that depends on both temperature and wavelength. Within
this model, the emissivity spectral index is therefore predicted to
change with both wavelength and temperature. The behaviour of a flattening of the spectrum in the millimeter range is in
qualitative agreement with their model, although the data do
not allow us to test further a possible dependence with dust temperature,
since the temperature range that we can sample here is limited.  Finally,
we note that in this model, the TLS phenomenon responsible for the
emission above $\rm \lambda \simeq 600\mic$ is of a different nature
than the vibration of DCD in the amorphous lattice producing the FIR
emission. It is therefore possible that fractal aggregates consisting of
amorphous individual grains also exhibit a change in properties
between the FIR and the submm. In particular, the TLS phenomenon
operating at atomic scale may be less sensitive to the dust
grains being gathered into aggregates, while the DCD vibration, which is
a global phenomenon, is expected to exhibit an excess emission, as
predicted by classical DDA calculations. This may offer a possible
physical interpretation to the emissivity increase of dust in the
molecular phase as being limited to the FIR region. Obviously, detailed
modelling of the interaction between the electromagnetic wave and a
fractal aggregate consisting of amorphous material is needed to further
investigate this issue.

\section{Conclusions}
\label{sec:conclusions}

We have analysed the dust emission from the outer Galactic plane
region using DIRBE, Archeops, and WMAP data from 100 $\mic$ to 3.2
mm. We have performed a correlation study of the FIR-mm emission with gas
tracers in individual regions, and derived the average equilibrium
temperature of large dust grains in both molecular and atomic
phases in a set of regions along the Galactic plane. We used this
temperature to derive the emissivity spectra for each phase and
region.  We classified regions into 3 classes, according to the relative
temperature of the dust associated with the molecular and atomic phases.
For each class, we derived the median emissivity profile.

We showed that the emissivity spectra are always steeper in the FIR
($\rm \lambda < 600\mic$) and flatten in the submm and mm. In regions
where dust is significantly colder in the molecular phase than in the
surrounding atomic medium, we produced an increase in the emissivity
by a factor of $\rm \simeq 3$ in the FIR. However, we showed that the
emissivity increase is restricted to the FIR range; the emissivity
spectra for the dust in the atomic and molecular phases become
comparable again in the submm and mm wavelength range.

The observed break in the emissivity spectrum, which appears to be a
general phenomenon, is consistent with the phenomenological all-sky
fit of the FIRAS data of \cite{Finkbeiner99}. It is also in 
qualitative agreement with the dust emission model of
\cite{Meny07}, which invokes quantum effects in amorphous solids to
explain the flatness of the observed submm emission spectrum and also
produces a break in the emissivity slope around 600 $\mic$.

We interpret the FIR emissivity increase in molecular clouds
containing cold dust as being caused by the coagulation of large grains into fractal
aggregates. Previous observations, obtained over much smaller
portions of the sky by \citet{Stepnik03}, showed that dust
aggregation could lead to an increase in the dust emissivity by a
factor of 3-4, to explain the unusually low dust temperatures
observed. The fact that emissivities do reconcile in the
submillimeter is not in agreement with DDA calculations of simple
aggregates and the physical reason for this remains unclear. We
however propose that this may be related to the amorphous nature of
the grains contained in the aggregates.

Finally, the absence of a detectable FIR emissivity increase in
regions where the dust temperature in the molecular phase is similar
or higher than that in the atomic phase is interpreted as the absence
of fractal grains in those environments. This is probably because the formation of such aggregates is prevented by the star
formation activity, while it is favoured in more quiescent regions
where turbulence is milder.

\section{Appendix}
\onecolumn
\begin{landscape}
\tablefirsthead{\hline \hline \multicolumn{1}{l}{\small Long.} &
\multicolumn{1}{l}{Lat.} &  pix. &\multicolumn{1}{l}{a$_{100}$} & \multicolumn{1}{l}{a$_{140}$} & \multicolumn{1}{l}{a$_{240}$} & \multicolumn{1}{l}{a$_{550}$} & \multicolumn{1}{l}{a$_{850}$} & \multicolumn{1}{l}{a$_{1382}$} & \multicolumn{1}{l}{a$_{2098}$} & \multicolumn{1}{l}{a$_{3191}$} & \multicolumn{1}{l}{$\rm \beta_{HI}^{FIR}$} & \multicolumn{1}{l}{$\rm \beta_{HI}^{submm}$} & \multicolumn{1}{l}{T$\rm _d^{HI}$} \\ 
 && &\multicolumn{1}{l}{$\pm\Delta$a$_{100}$}&\multicolumn{1}{l}{$\pm\Delta$a$_{140}$}&\multicolumn{1}{l}{$\pm \Delta$a$_{240}$}&\multicolumn{1}{l}{$\pm \Delta$a$_{550}$}&\multicolumn{1}{l}{$ \pm \Delta$a$_{850}$} &\multicolumn{1}{l}{$\pm \Delta$a$_{1382}$} & \multicolumn{1}{l}{$\pm \Delta$a$_{2098}$} & \multicolumn{1}{l}{$ \pm \Delta$a$_{3191}$} & \multicolumn{1}{l}{$\rm \pm \Delta \beta_{HI}^{FIR}$} &\multicolumn{1}{l}{$\rm \pm \Delta \beta_{HI}^{submm}$} & \multicolumn{1}{l}{$\pm \Delta $T$\rm _d^{HI}$} \\
& & && & & & ($\times 10^{-1}$) &  ($\times 10^{-2}$) &  ($\times 10^{-2}$) &  ($\times 10^{-3}$) & & & \\\hline}
\tablehead{\hline \multicolumn{14}{l}{\small\sl continued from previous page}
\\ \hline \multicolumn{1}{l}{\small Long.} &
\multicolumn{1}{l}{Lat.} & pix. &\multicolumn{1}{l}{a$_{100}$} & \multicolumn{1}{l}{a$_{140}$} & \multicolumn{1}{l}{a$_{240}$} & \multicolumn{1}{l}{a$_{550}$} & \multicolumn{1}{l}{a$_{850}$} & \multicolumn{1}{l}{a$_{1382}$} & \multicolumn{1}{l}{a$_{2098}$} & \multicolumn{1}{l}{a$_{3191}$} & \multicolumn{1}{l}{$\rm \beta_{HI}^{FIR}$} & \multicolumn{1}{l}{$\rm \beta_{HI}^{submm}$} & \multicolumn{1}{l}{T$\rm _d^{HI}$} \\ 
 &&&\multicolumn{1}{l}{$\pm\Delta$a$_{100}$}&\multicolumn{1}{l}{$\pm\Delta$a$_{140}$}&\multicolumn{1}{l}{$\pm \Delta$a$_{240}$}&\multicolumn{1}{l}{$\pm \Delta$a$_{550}$}&\multicolumn{1}{l}{$ \pm \Delta$a$_{850}$} &\multicolumn{1}{l}{$\pm \Delta$a$_{1382}$} & \multicolumn{1}{l}{$\pm \Delta$a$_{2098}$} & \multicolumn{1}{l}{$ \pm \Delta$a$_{3191}$} & \multicolumn{1}{l}{$\rm  \pm \Delta \beta_{HI}^{FIR}$} &\multicolumn{1}{l}{$\rm  \pm \Delta \beta_{HI}^{submm}$} & \multicolumn{1}{l}{$\pm \Delta $T$\rm _d^{HI}$} \\
 && & & & & & ($\times 10^{-1}$) &  ($\times 10^{-2}$) &  ($\times 10^{-2}$) &  ($\times 10^{-3}$) & & & \\  \hline}
\tabletail{\hline\multicolumn{14}{r}{\small\sl continued on next page}\\\hline } 
\tablelasttail{\hline}
\bottomcaption{Results of the IR/HI correlations in the atomic phase for all sky regions.
Cols 1-2: central position of each studied regions in degree. Col 3: number of pixels used for the correlations in each studied regions. Cols 4-11:
results of the correlations in $\rm {MJy/sr}$ for $\rm 10^{21}H/cm^{2}$. Cols 12-13: emissivity
spectral indices in the range 100 - 240 $\rm \mic$ ($\rm \beta^{FIR}$)
and in the range 550 - 2100 $\rm \mic$ ($\rm \beta^{submm}$). Col 14: derived dust temperature in Kelvin. Errors bars represent the 1-$\rm \sigma$ uncertainty estimates.
}
\begin{supertabular}{p{0.95cm}p{0.91cm}p{0.3cm}p{1.6cm}p{1.6cm}p{1.6cm}p{1.5cm}p{1.5cm}p{1.5cm}p{1.6cm}p{1.6cm}p{1.32cm}p{1.3cm}p{1.5cm}}
\hline
\hline
 Case 1  &&&&&&&&&&&&\\
\hline
108.34 &   8.28 &    71 &   12.0 $\pm$   1.3 &   26.4 $\pm$   2.3 &   18.9 $\pm$   1.4 &    3.2 $\pm$   0.2 &    9.6 $\pm$   0.4 &   20.3 $\pm$   0.8 &    6.1 $\pm$   0.4 &  13.5 $\pm$   1.9 &    2.3 $\pm$    0.5 &    1.6 $\pm$   0.1 &   17.2 $\pm$   0.7 \\
117.55 &  -2.16 &   111 &    9.4 $\pm$   0.8 &   20.0 $\pm$   1.6 &   15.6 $\pm$   1.1 &    2.1 $\pm$   0.2 &    8.1 $\pm$   0.4 &   14.3 $\pm$   0.7 &    4.3 $\pm$   0.3 &  10.1 $\pm$   1.1 &    2.3 $\pm$    0.5 &    1.6 $\pm$   0.1 &   16.6 $\pm$   0.6 \\
120.61 &  -6.34 &    40 &    7.1 $\pm$   6.5 &   16.0 $\pm$   8.3 &   13.9 $\pm$   4.3 &    2.3 $\pm$   0.4 &    7.5 $\pm$   1.0 &   17.6 $\pm$   2.8 &    4.3 $\pm$   1.3 &  22.7 $\pm$   6.2 &    2.2 $\pm$   1.9 &    1.6 $\pm$   0.3 &   17.0 $\pm$   0.6 \\
139.02 &  12.46 &    17 &    2.1 $\pm$   1.6 &    6.6 $\pm$   4.1 &    7.1 $\pm$   4.7 &    1.9 $\pm$   0.5 &    2.4 $\pm$   3.1 &    8.6 $\pm$   3.9 &    1.0 $\pm$   2.0 &  11.1 $\pm$  12.5 &    2.0 $\pm$   3.1 &    2.1 $\pm$   0.7 &   15.7 $\pm$   0.6 \\
142.09 &   2.01 &   111 &    5.3 $\pm$   1.1 &   13.7 $\pm$   1.8 &   12.1 $\pm$   1.0 &    1.7 $\pm$   0.1 &    7.9 $\pm$   0.4 &   12.4 $\pm$   0.6 &    3.6 $\pm$   0.3 &  10.7 $\pm$   1.0 &    2.4 $\pm$   0.6 &    1.7 $\pm$   0.1 &   15.3$\pm$    0.5 \\
145.16 &  -4.25 &    91 &    4.1 $\pm$   0.2 &   10.9 $\pm$   0.5 &    9.2 $\pm$   0.4 &    1.5  $\pm$  0.1 &    6.0 $\pm$   0.3 &    9.9 $\pm$   0.5 &    3.5 $\pm$   0.3 &  $6.6$ $\pm$    1.1 &    2.3 $\pm$    0.4 &    1.6 $\pm$   0.1 &   15.5 $\pm$    0.5 \\
148.23 &  -6.34 &    38 &    4.1 $\pm$   0.3 &   10.2 $\pm$   1.0 &    8.4 $\pm$   0.7 &    0.2 $\pm$   0.1 &    4.8 $\pm$   0.6 &   11.9 $\pm$   1.0 &    4.8 $\pm$   0.5 &  $6.2$ $\pm$   1.7 &    5.2 $\pm$   0.4 &    1.0 $\pm$   0.2 &   10.6 $\pm$   0.2 \\
148.23 &   2.01 &   111 &    3.9 $\pm$   0.3 &   10.6 $\pm$   0.6 &   10.1 $\pm$   0.5 &    1.4 $\pm$   0.1 &    7.8 $\pm$   0.3 &   11.8 $\pm$   0.5 &    3.7 $\pm$   0.2 &  10.2 $\pm$   0.7 &    2.4 $\pm$   0.4 &    1.6 $\pm$   0.1 &   14.7 $\pm$   0.5 \\
151.29 &  -0.07 &   110 &    9.9 $\pm$   2.1 &   17.0 $\pm$   3.3 &   12.8 $\pm$   2.2 &    1.4 $\pm$   0.4 &    6.8 $\pm$   1.1 &   22.7 $\pm$   2.0 &    9.5 $\pm$   1.0 &  28.0 $\pm$    3.2 &    2.3 $\pm$   0.6 &    0.6 $\pm$   0.2 &   17.7 $\pm$   0.6 \\
151.29 &   6.19 &   100 &    2.7 $\pm$   0.3 &    6.3 $\pm$   0.8 &    5.6 $\pm$   0.7 &    0.9 $\pm$   0.2 &    5.2 $\pm$   0.5 &    7.8 $\pm$   0.7 &    2.4 $\pm$   0.4 &  $9.6$ $\pm$    2.4 &    2.1 $\pm$   0.5 &    1.7 $\pm$   0.2 &   16.5 $\pm$   0.6 \\
154.36 & -10.52 &    51 &    1.8 $\pm$   0.4 &    6.2 $\pm$   1.4 &    6.1 $\pm$   1.4 &    1.1 $\pm$   0.2 &    2.6 $\pm$   0.8 &   - &   - & $<$1.72 &    2.4 $\pm$   0.8 &    2.5 $\pm$   1.1 &   13.9 $\pm$   0.5 \\
157.43 &  12.46 &    33 &    4.1 $\pm$   0.8 &   12.5 $\pm$   3.0 &    7.9 $\pm$   2.5 &    3.2 $\pm$   0.7 &    5.4 $\pm$   1.5 &    7.9 $\pm$   2.9 &  $<$ 5.2 & 17.0 $\pm$   9.1 &    3.1 $\pm$   1.1 &    2.9 $\pm$   0.6 &   13.9 $\pm$   0.7 \\
160.50 &  -8.43 &   112 &    6.8 $\pm$   1.6 &   19.5 $\pm$   4.7 &   20.8 $\pm$   4.4 &    1.8 $\pm$   0.6 &    5.3 $\pm$   2.3 &   16.6 $\pm$   4.9 &    5.2 $\pm$   2.4 &  23.3 $\pm$   10.5 &    2.9 $\pm$    0.7 &    1.4 $\pm$   0.4 &   13.0 $\pm$   0.3 \\
160.50 &   2.01 &   117 &    2.9 $\pm$   0.2 &    9.6 $\pm$   0.6 &    9.9 $\pm$   0.5 &    2.4 $\pm$   0.1 &    5.0 $\pm$   0.5 &   12.8 $\pm$   0.9 &    5.2 $\pm$   0.4 &  $6.1$ $\pm$    1.8 &    2.0 $\pm$    0.5 &    1.5 $\pm$   0.1 &   15.0 $\pm$   0.5 \\
163.57 &   8.28 &    42 &    6.8 $\pm$   0.8 &   15.8 $\pm$   2.6 &   15.0 $\pm$   2.2 &    0.9 $\pm$   0.3 &    6.4 $\pm$   1.2 &    8.8 $\pm$   3.9 &    3.4 $\pm$   2.5 &  $9.6$ $\pm$  11.2 &    3.2 $\pm$    0.5 &    1.3 $\pm$   0.6 &   13.2 $\pm$   0.3 \\
163.57 &  10.37 &    59 &    5.4 $\pm$   0.5 &   14.8 $\pm$   1.6 &   13.5 $\pm$   1.2 &    0.6 $\pm$   0.2 &    4.2 $\pm$   0.9 &    2.2 $\pm$   2.5 &   $<$ 2.7 &  $3.6$ $\pm$   7.3 &    4.0 $\pm$   0.5 &    0.2 $\pm$   1.2 &   11.8 $\pm$    0.3 \\
166.63 &   8.28 &    34 &    6.8 $\pm$   0.9 &   16.6 $\pm$   3.0 &   16.0 $\pm$   2.6 &    1.3 $\pm$   0.3 &    6.7 $\pm$   1.4 &    9.1 $\pm$   3.7 &    2.7 $\pm$ 1.9&   $<$30.4 &    2.8 $\pm$   0.6 &    1.6 $\pm$   0.5 &   13.9 $\pm$   0.4 \\
172.77 & -10.52 &   110 &    3.2 $\pm$   0.4 &   10.3 $\pm$   1.4 &    8.2 $\pm$   1.3 &    0.8 $\pm$   0.3 &   $<$ 1.9 &   10.3 $\pm$   1.2 &    0.9 $\pm$   0.8 &- &    3.3 $\pm$   0.6 &    1.2 $\pm$   0.6 &   12.9 $\pm$   0.4 \\
\hline
Median & values &&&&&&&&&&&&\\
 - & - & - &    5.3 $\pm$   0.8 &   13.7 $\pm$   1.8 &   12.1 $\pm$  1.4 &    1.5 $\pm$   0.2 &    6.0 $\pm$   0.9 &   11.8 $\pm$   1.5 &    3.6 $\pm$   0.8 &  $9.6$ $\pm$   3.2 &    2.4 $\pm$   0.6 &    1.6 $\pm$   0.3 &   15.0 $\pm$   0.5 \\

\hline
\hline
 Case 2  &&&&&&&&&&&&\\
\hline
 71.53 &   4.10 &    64 &    4.5$\pm$    0.9 &    8.5$\pm$    1.8 &    7.3$\pm$    1.3 &    1.7$\pm$    0.1 &    7.8 $\pm$   0.6 &   10.6  $\pm$  0.6 &    3.3 $\pm$   0.3 &  $7.9$  $\pm$  1.2 &    1.7  $\pm$  0.7 &    1.7 $\pm$   0.1 &   20.7 $\pm$   0.9 \\
 74.59 &   2.01 &    86 &   10.4$\pm$    3.7 &   20.0$\pm$    4.5 &   14.4$\pm$    2.0 &    2.5$\pm$    0.2 &    7.6 $\pm$   0.9 &   14.8 $\pm$   1.1 &    4.6 $\pm$   0.5 &  $9.6$ $\pm$   1.3 &    2.2 $\pm$  0.8 &    1.6  $\pm$  0.2 &   18.8 $\pm$   0.8 \\
 77.66 &   4.10 &   104 &   10.4$\pm$    3.0 &   17.5$\pm$    3.7 &   12.0$\pm$    1.5 &    2.1$\pm$    0.2 &    6.3 $\pm$   0.4 &   13.3 $\pm$   0.8 &    3.5 $\pm$   0.3 &  $6.8$ $\pm$   1.3 &    2.1  $\pm$  0.7 &    1.7 $\pm$   0.1 &   20.7 $\pm$   0.9 \\
 80.73 &   8.28 &    56 &   17.2$\pm$    1.2 &   30.2$\pm$    1.7 &   18.6$\pm$    0.9 &    2.9$\pm$    0.1 &    8.0  $\pm$  0.3 &   16.7  $\pm$  0.5 &    6.0 $\pm$   0.4 &  $8.9$  $\pm$  1.8 &    2.3 $\pm$   0.4 &    1.6 $\pm$   0.1 &   19.5  $\pm$  0.8 \\
 83.80 &  -4.25 &    59 &   10.4$\pm$    2.5 &   21.3$\pm$    4.5 &   14.5$\pm$    3.0 &    3.2$\pm$    0.4 &   10.7 $\pm$   1.6 &   12.2 $\pm$   2.7 &    6.0 $\pm$   1.1 &  19.6  $\pm$  4.1 &    2.1  $\pm$  0.8 &    1.7  $\pm$  0.2 &   19.2  $\pm$  0.9 \\
 89.93 &   2.01 &   111 &   16.3$\pm$    1.4 &   28.9$\pm$   2.1 &   19.0 $\pm$   1.3 &    2.7$\pm$    0.2 &   10.1$\pm$    0.4 &   22.0  $\pm$  0.9 &    6.8  $\pm$  0.3 &  16.9 $\pm$   1.5 &    2.3   $\pm$ 0.5 &    1.4  $\pm$  0.1 &   18.9 $\pm$   0.8 \\
 93.00 &  -0.07 &   111 &    5.7$\pm$    0.6 &   12.3$\pm$    1.1 &    9.8$\pm$    0.8 &    2.0$\pm$    0.1 &    6.9 $\pm$   0.3 &   14.1 $\pm$   0.6 &    3.9 $\pm$   0.3 &  10.1 $\pm$   0.9 &    2.0 $\pm$   0.5 &    1.6  $\pm$  0.1 &   18.3  $\pm$  0.7 \\
 99.14 &  -4.25 &    91 &    6.4$\pm$    0.3 &   15.1$\pm$   0.6 &   12.6$\pm$    0.5 &    1.8$\pm$    0.2 &    8.0  $\pm$  0.5 &   17.4  $\pm$  0.6 &    5.8  $\pm$  0.4 &  $4.5$ $\pm$   1.7 &    2.3 $\pm$   0.4 &    1.3 $\pm$   0.1 &   15.9 $\pm$   0.5 \\
 99.14 &  -0.07 &   110 &    5.8$\pm$    0.7 &   15.0 $\pm$   1.0 &   13.5 $\pm$   0.7 &    3.0$\pm$    0.1 &    9.1 $\pm$   0.3 &   18.2 $\pm$   0.6 &    4.3 $\pm$   0.3 &  $7.0$ $\pm$   1.4 &    2.0 $\pm$   0.5 &    1.8 $\pm$   0.1 &   16.8  $\pm$  0.6 \\
 99.14 &  12.46 &    79 &   11.1$\pm$    0.4 &   25.0$\pm$    1.0 &   19.4$\pm$    0.8 &    2.8$\pm$    0.1 &   10.5 $\pm$   0.4 &   20.7 $\pm$   1.0 &    6.6 $\pm$   0.6 &  $6.7$ $\pm$   4.2 &    2.3 $\pm$   0.4 &    1.5  $\pm$  0.1 &   16.4  $\pm$  0.6 \\
102.21 &  10.37 &    99 &    9.7$\pm$    1.7 &   21.1$\pm$    4.0 &   14.7$\pm$    3.2 &    2.0$\pm$    0.4 &    9.8 $\pm$   1.3 &   11.8  $\pm$  2.4 &    4.1 $\pm$   1.2 & $<$8.7 &    2.5  $\pm$  0.7 &    1.7 $\pm$   0.3 &   16.5 $\pm$   0.6 \\
102.21 &  12.46 &    68 &   12.3$\pm$    0.5 &   26.8$\pm$    1.1 &   21.3 $\pm$   0.8 &    2.8$\pm$    0.1 &   11.9 $\pm$   0.4 &   19.9 $\pm$   0.7 &    6.8 $\pm$   0.6 &  $7.2$ $\pm$   2.5 &    2.3  $\pm$  0.4 &    1.6 $\pm$   0.1 &   16.4  $\pm$   0.6 \\
108.34 &  -4.25 &    90 &   12.0$\pm$    0.6 &   25.1$\pm$    1.2 &   18.1$\pm$    0.8 &    2.7$\pm$    0.3 &    9.0  $\pm$  0.4 &   17.6 $\pm$   0.5 &    5.7 $\pm$   0.2 &  $9.4$ $\pm$   1.4 &    2.3 $\pm$   0.4 &    1.5 $\pm$   0.1 &   17.2  $\pm$   0.6 \\
108.34 &   2.01 &   111 &   21.8$\pm$    5.1 &   26.0$\pm$    6.5 &   13.9$\pm$    3.6 &    1.1$\pm$    0.5 &    5.8 $\pm$   1.0 &    9.5 $\pm$   2.2 &    2.8 $\pm$   0.7 & $<$3.7 &    2.7 $\pm$   0.8 &    1.6 $\pm$   0.4 &   20.8  $\pm$   0.8 \\
111.41 &   6.19 &    83 &    2.0$\pm$    1.4 &    5.3$\pm$    2.4 &    6.1$\pm$    1.4 &    1.7$\pm$    0.2 &    3.4 $\pm$   0.4 &    8.4 $\pm$   0.6 &    2.2 $\pm$   0.2 &  $5.5$ $\pm$   0.8 &    1.6 $\pm$   1.7 &    1.8 $\pm$   0.2 &   17.4 $\pm$    0.6 \\
114.48 &  -0.07 &   110 &    6.1$\pm$    4.1 &   14.3$\pm$    6.1 &   11.4$\pm$    3.5 &    2.2$\pm$    0.4 &    8.6 $\pm$   1.0 &   18.1 $\pm$   1.7 &    4.1 $\pm$   0.4 & $<$6.8 &    2.2 $\pm$   1.5 &    1.7 $\pm$   0.2 &   17.3  $\pm$   0.7 \\
117.55 &  -0.07 &   110 &    4.4$\pm$    2.1 &    9.6$\pm$    4.1 &    8.9$\pm$    2.7 &    1.0 $\pm$   0.4 &    6.3 $\pm$   1.0 &   13.6 $\pm$   1.5 &    2.8 $\pm$   0.4 &  $9.3$  $\pm$  2.1 &    2.3 $\pm$   1.1 &    1.7 $\pm$   0.3 &   15.6  $\pm$   0.5 \\
123.68 &   2.01 &   111 &    5.2$\pm$    0.4 &   11.8$\pm$    0.8 &   10.6$\pm$    0.7 &    1.5$\pm$    0.1 &    6.6 $\pm$   0.3 &   13.5 $\pm$   0.6 &    3.8 $\pm$   0.3 &  12.6   $\pm$   1.2 &    2.2 $\pm$   0.4 &    1.5 $\pm$   0.1 &   16.4 $\pm$    0.5 \\
129.82 &   2.01 &   110 &    4.9$\pm$    0.3 &   11.7$\pm$    0.7 &   10.3$\pm$    0.4 &    1.9$\pm$    0.1 &    5.3 $\pm$   0.2 &   13.7  $\pm$  0.4 &    3.3 $\pm$   0.2 &  $6.6$ $\pm$   1.1 &    2.0 $\pm$   0.4 &    1.6 $\pm$   0.1 &   16.9  $\pm$   0.6 \\
135.95 &  -4.25 &    49 &    6.4$\pm$    0.4 &   14.1$\pm$    0.9 &   12.4$\pm$    0.8 &    1.6$\pm$    0.1 &    8.1 $\pm$   0.4 &   12.0 $\pm$   0.9 &    4.4 $\pm$   0.6 &  $3.2$ $\pm$   3.3 &    2.2  $\pm$  0.4 &    1.5  $\pm$  0.2 &   16.3  $\pm$   0.5 \\
135.95 &  -2.16 &   106 &    4.6$\pm$   0.5 &   10.9$\pm$    1.0 &    9.7$\pm$    0.7 &    2.1 $\pm$   0.1 &    5.0  $\pm$  0.4 &   12.8  $\pm$  0.7 &    2.7  $\pm$  0.3 &  10.5   $\pm$   1.1 &    1.9  $\pm$  0.5 &    1.8  $\pm$  0.1 &   17.5 $\pm$    0.7 \\
139.02 &  -4.25 &    62 &    6.1$\pm$    0.4 &   13.7$\pm$    1.0 &   11.4$\pm$    0.9 &    2.1$\pm$    0.2 &    7.9 $\pm$   0.5 &   12.6 $\pm$   0.9 &    4.0 $\pm$   0.4 &  $4.3$ $\pm$   1.8 &    2.0  $\pm$  0.5 &    1.7 $\pm$   0.2 &   17.4 $\pm$    0.7 \\
139.02 &  -0.07 &   111 &   23.1$\pm$    4.1 &   39.4$\pm$    5.7 &   25.4$\pm$    3.0 &    3.1$\pm$    0.4 &   14.4 $\pm$   1.1 &   21.8$\pm$    1.6 &    4.9 $\pm$   0.6 &  $9.1$$\pm$    1.7 &    2.4 $\pm$   0.6 &    1.9 $\pm$   0.2 &   18.6  $\pm$   0.7 \\
139.02 &   2.01 &   110 &   18.5$\pm$    2.6 &   29.7$\pm$    3.6 &   18.8$\pm$    1.9 &    2.4$\pm$    0.2 &    8.5  $\pm$  0.6 &   14.5 $\pm$   0.8 &    4.6 $\pm$   0.3 & 13.1  $\pm$     1.2 &    2.4 $\pm$   0.5 &    1.6 $\pm$   0.1 &   19.6  $\pm$   0.8 \\
145.16 &   4.10 &   111 &    6.7$\pm$    0.5 &   14.6$\pm$    0.9 &   11.7$\pm$    0.6 &    2.8$\pm$    0.2 &    9.4 $\pm$   0.3 &   16.4 $\pm$   1.1 &    4.5 $\pm$   0.4 &  $5.4$ $\pm$   2.7 &    1.9 $\pm$   0.5 &    1.7 $\pm$   0.1 &   18.8  $\pm$   0.8 \\
145.16 &  10.37 &   116 &    4.8$\pm$    0.2 &   11.6$\pm$    0.8 &   11.1$\pm$    0.7 &    1.9$\pm$    0.2 &    6.3 $\pm$   0.4 &   16.1 $\pm$   1.1 &    4.8 $\pm$   0.6 &  11.4  $\pm$  2.5 &    2.0  $\pm$  0.4 &    1.3 $\pm$   0.2 &   16.3 $\pm$    0.6 \\
151.29 &   4.10 &   110 &    4.3$\pm$    0.3 &   10.0$\pm$    0.7 &    9.6$\pm$    0.6 &    1.9$\pm$    0.1 &    7.8 $\pm$   0.3 &   10.8 $\pm$   0.7 &    3.7  $\pm$  0.3 &  $8.2$  $\pm$  1.5 &    1.9  $\pm$  0.4 &    1.7  $\pm$  0.1 &   17.1 $\pm$    0.6 \\
154.36 &   4.10 &   111 &    3.6$\pm$    0.4 &    7.6$\pm$    0.8 &    7.0$\pm$    0.6 &    1.3$\pm$    0.1 &    7.0 $\pm$   0.2 &    7.9 $\pm$   0.6 &    1.6 $\pm$   0.3 &  $2.1$$\pm$    0.8 &    1.9 $\pm$   0.5 &    2.1  $\pm$  0.2 &   18.0  $\pm$   0.7 \\
157.43 &  -0.07 &   110 &    5.1$\pm$    0.3 &   12.6$\pm$    0.9 &   11.4$\pm$    0.9 &    2.5$\pm$    0.2 &    5.6   $\pm$ 0.9 &   13.6$\pm$    1.7 &    2.1 $\pm$   0.7 &  $2.2$ $\pm$   2.4 &    1.9 $\pm$   0.5 &    2.0 $\pm$   0.2 &   17.2   $\pm$  0.6 \\
160.50 & -20.96 &    72 &    6.6$\pm$    6.4 &   14.6$\pm$   13.6 &    8.7 $\pm$   8.9 &    3.8$\pm$    1.3 &    9.5$\pm$    5.4 &   20.2 $\pm$   8.8 &    6.4  $\pm$  4.0 & $<$ 15.6 & - &    1.8 $\pm$   0.6 &   16.8  $\pm$   1.1 \\
160.50 & -18.87 &    91 &   24.7$\pm$    8.6 &   50.5$\pm$   15.6 &   31.8$\pm$    9.7 &    3.5$\pm$    0.9 &   10.4$\pm$    4.1 &   35.9 $\pm$   7.4 &   17.1 $\pm$   1.9 &  65.1 $\pm$  5.9 &    2.7 $\pm$   1.0 &    0.8  $\pm$  0.3 &   16.3 $\pm$    0.6 \\
160.50 &  -0.07 &   117 &    4.1$\pm$    0.5 &   10.4$\pm$    1.4 &   10.0$\pm$    1.3 &    3.0 $\pm$   0.2 &   11.3 $\pm$   1.0 &   18.4 $\pm$   2.1 &    4.8 $\pm$   0.8 &  15.0 $\pm$   2.6 &    1.6 $\pm$   0.6 &    1.7  $\pm$  0.2 &   18.2 $\pm$    0.8 \\
169.70 &  -6.34 &   110 &    3.1$\pm$    0.4 &    8.1$\pm$    1.3 &    6.8$\pm$    1.2 &    0.2$\pm$    0.2 &    3.1 $\pm$   0.7 &   $<$2.3 &  $<$1.5 & $<$7.3  &    4.4 $\pm$   0.6 & - &   11.5 $\pm$    0.3 \\
172.77 & -18.87 &   111 &    6.5$\pm$    0.3 &   14.8$\pm$    1.1 &   11.9$\pm$    1.0 &    2.4$\pm$    0.2 &    4.4  $\pm$  0.6 &   11.4$\pm$    0.8 &    3.4 $\pm$   0.4 &  $4.7$ $\pm$   2.0 &    2.0 $\pm$   0.5 &    1.8$\pm$    0.2 &   17.7 $\pm$    0.7 \\
172.77 & -14.69 &   110 &    9.5$\pm$    0.7 &   28.5$\pm$    2.4 &   29.5$\pm$    3.1 &    6.1$\pm$    0.5 &   18.8  $\pm$  1.7 &   28.7 $\pm$   2.5 &    7.6 $\pm$   0.9 &  $8.5$ $\pm$   3.2 &    2.0  $\pm$  0.5 &    2.0  $\pm$  0.2 &   15.3  $\pm$   0.5 \\
172.77 &  -8.43 &    91 &    6.5$\pm$    0.5 &   18.2$\pm$    1.7 &   15.9$\pm$    1.5 &    3.2$\pm$    0.4 &    6.9 $\pm$   1.5 &   16.9 $\pm$   1.8 &   10.3 $\pm$   1.3 &  29.0 $\pm$   6.7 &    2.2  $\pm$  0.5 &    1.3  $\pm$  0.2 &   15.8  $\pm$   0.6 \\
172.77 &  -6.34 &    81 &    2.2$\pm$    0.4 &    5.3$\pm$    1.4 &    4.7$\pm$    1.2 &    0.7 $\pm$   0.2 &    0.8 $\pm$   0.7 &   $<$2.2 &   $<$1.7 & $<$2.2 &    2.2  $\pm$  0.7 &    3.9 $\pm$   3.7 &   15.8  $\pm$   0.5 \\
175.84 &   4.10 &   109 &    5.7$\pm$    1.6 &   13.4$\pm$    2.6 &   11.8$\pm$    1.6 &    1.4 $\pm$   0.2 &    9.1 $\pm$   0.6 &   15.0 $\pm$   1.2 &    2.3 $\pm$   0.7 &  0.2 $\pm$   3.0 &    2.4  $\pm$  0.7 &    1.6  $\pm$  0.2 &   15.4 $\pm$    0.5 \\
178.91 & -23.05 &    51 &    8.6$\pm$    2.0 &   18.8$\pm$    7.2 &   23.3$\pm$    5.8 &    4.4$\pm$    1.4 &   18.7 $\pm$   2.1 &   17.2 $\pm$   5.0 &    1.7 $\pm$   3.8 &  $9.8$ $\pm$  20.1 &    1.5 $\pm$   0.8 &    2.4 $\pm$   0.6 &   17.5  $\pm$   0.6 \\
178.91 &   2.01 &   110 &    6.3$\pm$    0.5 &   18.4$\pm$    1.4 &   19.0$\pm$    1.2 &    5.4 $\pm$   0.5 &   11.3 $\pm$   0.9 &   25.3 $\pm$   1.7 &    8.6 $\pm$   0.5 & 10.6 $\pm$    1.5 &    1.7  $\pm$  0.5 &    1.6  $\pm$  0.1 &   16.5$\pm$     0.6 \\
181.97 & -12.61 &    66 &    3.7$\pm$    1.0 &   10.9$\pm$    3.3 &    8.7$\pm$    2.1 &    2.3   $\pm$ 0.4 &    8.6 $\pm$   0.6 &    7.3  $\pm$  2.2 &   $<$3.9 & $<$3.2 &    2.2  $\pm$  0.9 &    2.2 $\pm$   0.5 &   15.9  $\pm$   0.7 \\
181.97 &  -8.43 &    48 &    2.2$\pm$    1.2 &   11.0$\pm$    3.3 &   14.7$\pm$    2.8 &    1.2 $\pm$   0.5 &    8.5 $\pm$   0.9 &   19.8 $\pm$   2.8 &    4.1 $\pm$   1.2 &  $3.2$ $\pm$   5.5 &    4.3 $\pm$   1.0 &    1.7 $\pm$   0.3 &   10.4 $\pm$    0.2 \\
181.97 &  -2.16 &   111 &    5.1$\pm$    0.6 &   14.8$\pm$    1.8 &   16.4$\pm$    1.6 &    3.2 $\pm$   0.5 &   11.3  $\pm$  0.8 &   24.8 $\pm$   1.4 &    7.8  $\pm$  0.6 &  27.2 $\pm$    2.9 &    1.9 $\pm$   0.5 &    1.4 $\pm$   0.2 &   15.3  $\pm$   0.5 \\
181.97 &   2.01 &   111 &    7.9$\pm$    0.5 &   20.6$\pm$    1.1 &   19.0$\pm$    0.9 &    4.1 $\pm$   0.3 &   12.6 $\pm$   0.4 &   26.4 $\pm$   0.9 &    7.8 $\pm$   0.5 &  $7.5$$\pm$    2.1 &    2.0 $\pm$   0.4 &    1.6 $\pm$   0.1 &   16.6 $\pm$    0.6 \\
185.04 & -10.52 &    48 &    5.0$\pm$    0.6 &   13.2$\pm$    2.6 &   12.4$\pm$    1.9 &    0.8 $\pm$   0.2 &   10.4  $\pm$  0.8 &   13.2  $\pm$  1.9 &    1.4  $\pm$  0.9 & - &    3.3  $\pm$  0.6 &    1.9 $\pm$   0.3 &   12.9  $\pm$   0.3 \\

\hline
Median & values  &&&&&&&&&&&&\\
 - &   - &    - &    6.3$\pm$    0.6 &   14.6 $\pm$   1.7 &   12.4$\pm$    1.3 &    2.3 $\pm$   0.2 &    8.5 $\pm$  0.6 &   14.8$\pm$    1.1 &    4.1 $\pm$   0.5 &  $7.2$$\pm$    2.1 &    2.2 $\pm$  0.5 &    1.7 $\pm$    0.2 &   16.9 $\pm$   0.6 \\
\hline
\hline
 Case 3  &&&&&&&&&&&&\\
\hline
 68.46 &   4.10 &    34 &    3.7 $\pm$   1.5 &   13.1 $\pm$    3.1 &   11.7 $\pm$    2.0 &    2.5  $\pm$   0.2 &    4.1 $\pm$    1.9 &    9.4  $\pm$   1.3 &    3.4  $\pm$   0.8 &  $5.2$  $\pm$   2.9 &    2.6  $\pm$   0.9 &    2.1 $\pm$    0.2 &   14.0  $\pm$   0.5 \\
 77.66 &  -0.07 &    61 &   38.6 $\pm$  10.7 &   50.4 $\pm$   12.0 &   23.8 $\pm$    5.0 &    1.1 $\pm$    0.4 &   19.3  $\pm$   2.5 &   16.5  $\pm$   2.9 &    7.3 $\pm$    1.4 &  $8.8$  $\pm$   2.0 &    3.3 $\pm$    0.7 &    1.7 $\pm$    0.3 &   17.0 $\pm$    0.6 \\
 77.66 &   2.01 &   107 &   10.3 $\pm$   7.2 &   25.4 $\pm$    8.2 &   19.0  $\pm$   3.3 &    2.8 $\pm$    0.3 &    9.6  $\pm$   1.4 &   17.9  $\pm$   1.8 &    5.8 $\pm$    0.9 &  $8.6$  $\pm$   1.5 &    2.6  $\pm$   1.3 &    1.6$\pm$    0.2 &   15.8  $\pm$   0.6 \\
 83.80 &  -0.07 &   105 &    8.9  $\pm$  7.3 &   26.7 $\pm$  10.1 &   21.2   $\pm$  4.8 &    2.9 $\pm$   0.3 &   11.6  $\pm$   1.3 &   20.1 $\pm$    2.3 &    5.4 $\pm$    0.9 &  16.7  $\pm$   2.0 &    3.0  $\pm$   1.5 &    1.7   $\pm$  0.2 &   14.0  $\pm$  0.5 \\
 93.00 & -12.61 &    17 &    7.3  $\pm$  1.8 &   16.5 $\pm$    5.1 &   15.4 $\pm$    2.9 &    2.9 $\pm$    0.3 &    3.8  $\pm$   1.6 &   17.1 $\pm$    2.3 &    6.7  $\pm$   1.7 &  30.6 $\pm$    8.1 &    1.9 $\pm$    0.7 &    1.6 $\pm$    0.2 &   17.4  $\pm$   0.6 \\
 96.07 &  -6.34 &    43 &    2.9  $\pm$   8.1 &   10.1 $\pm$   10.5 &   11.3 $\pm$    4.8 &    2.7 $\pm$    0.4 &    7.6 $\pm$    0.8 &    9.9 $\pm$    1.8 &    1.1 $\pm$    1.3 &  $3.1$ $\pm$    5.8 & - &    2.4 $\pm$    0.4  &  14.4 $\pm$   0.5 \\
 96.07 &  -4.25 &   100 &    3.6  $\pm$   1.6 &   12.1 $\pm$    2.2 &   12.0 $\pm$    1.2 &    1.6  $\pm$   0.2 &    7.0  $\pm$   0.5 &   19.1 $\pm$    1.1 &    5.4 $\pm$    0.5 &  $8.5$ $\pm$    1.7 &    2.8  $\pm$   0.8 &    1.2 $\pm$   0.1 &   13.2 $\pm$    0.4 \\
102.21 &  -0.07 &   111 &    3.8  $\pm$   1.8 &   12.6 $\pm$    2.6 &   12.4  $\pm$   1.5 &    2.4 $\pm$    0.1 &    8.2  $\pm$   0.5 &   15.8 $\pm$    1.0 &    4.4 $\pm$    0.4 &  $9.0$  $\pm$   1.8 &    2.4  $\pm$   0.8 &    1.7   $\pm$  0.1 &   14.3 $\pm$    0.5 \\
105.27 &  -0.07 &   111 &    3.6 $\pm$    1.9 &   12.0 $\pm$    2.7 &   11.9 $\pm$    1.5 &    2.0 $\pm$    0.2 &    7.7  $\pm$   0.4 &   15.6 $\pm$    0.9 &    4.7 $\pm$    0.3 &  $9.1$ $\pm$    1.2 &    2.5  $\pm$   0.9 &    1.5  $\pm$   0.1 &   14.0  $\pm$   0.4 \\
126.75 &  -2.16 &   108 &    2.1 $\pm$    1.0 &    7.5 $\pm$    1.4 &    7.6 $\pm$    0.9 &    1.0 $\pm$    0.2 &    5.6  $\pm$   0.4 &   10.7  $\pm$   0.6 &    4.2  $\pm$   0.3 &  $7.6$  $\pm$   1.2 &    3.0  $\pm$   0.8 &    1.2 $\pm$   0.1 &   12.8 $\pm$    0.4 \\
142.09 &  -2.16 &   111 &    6.1  $\pm$   0.8 &   16.1 $\pm$    1.4 &   13.8 $\pm$    0.9 &    3.1 $\pm$    0.1 &    8.2 $\pm$    0.4 &   18.2 $\pm$    0.7 &    3.2  $\pm$   0.3 &  $7.1$  $\pm$   1.2 &    2.1  $\pm$   0.5 &   1.9 $\pm$    0.1 &   16.6  $\pm$   0.6 \\
142.09 &  12.46 &    55 &    2.4  $\pm$   0.7 &    7.7 $\pm$    2.0 &    9.9 $\pm$    1.8 &    3.1 $\pm$    0.4 &    7.8 $\pm$    2.2 &   24.0  $\pm$   3.4 &    6.2 $\pm$    1.2 &  17.7  $\pm$   5.5 & - &    2.0 $\pm$   0.7 &     8.2 $\pm$    1.4 \\
145.16 &  -2.16 &   110 &    4.1 $\pm$    0.4 &   13.3  $\pm$   0.8 &   12.4 $\pm$    0.7 &    2.9  $\pm$   0.1 &    7.4 $\pm$    0.4 &   17.2  $\pm$   0.8 &    3.7 $\pm$    0.3 &  13.6  $\pm$  0.9 &    2.2  $\pm$   0.5 &    1.9 $\pm$    0.1 &   15.0  $\pm$   0.6 \\
148.23 &  -2.16 &   111 &    4.0 $\pm$    0.7 &   12.8 $\pm$   1.2 &   11.7 $\pm$    1.0 &    3.4 $\pm$    0.2 &    6.8 $\pm$    0.6 &   16.4  $\pm$   1.1 &    3.8 $\pm$    0.4 &  16.2  $\pm$   1.6 &    2.1  $\pm$   0.6 &    1.9 $\pm$    0.1 &   15.7   $\pm$ 0.6 \\
154.36 &  -8.43 &    53 &    2.4  $\pm$   0.8 &    8.2 $\pm$    2.0 &    4.8  $\pm$   2.3 &    1.7 $\pm$    0.3 &    5.1  $\pm$   1.3 &   $<$4.8 &   $<$3.7 & $<$16.1   &  6.5 $\pm$   1.8  &   2.4 $\pm$    1.1  &   9.6 $\pm$    0.4 \\
154.36 &  -4.25 &    98 &    3.5 $\pm$    0.2 &    9.1  $\pm$   0.5 &    8.2  $\pm$   0.4 &    0.3 $\pm$    0.1 &    3.8 $\pm$    0.3 &   11.1  $\pm$   0.5 &    3.9 $\pm$    0.3 &  $9.7$  $\pm$   1.3 &    4.3  $\pm$  0.4 &    1.0 $\pm$    0.2 &   11.5  $\pm$   0.3 \\
154.36 &  14.54 &    20 &    8.8 $\pm$   3.2 &   34.4 $\pm$   14.9 &   17.2  $\pm$  11.4 &    0.9 $\pm$    2.8 &    1.6 $\pm$    7.5 &    7.2  $\pm$  13.5 &   28.7 $\pm$   12.8 &  25.5 $\pm$   44.8 &    6.1  $\pm$   2.1 & - &    9.9  $\pm$   0.3 \\
157.43 &   2.01 &   111 &    3.7  $\pm$   0.3 &   11.2 $\pm$    0.5 &   11.2 $\pm$    0.5 &    2.1 $\pm$    0.1 &    7.6 $\pm$    0.4 &   15.7 $\pm$    0.7 &    4.1 $\pm$    0.4 &  $1.6$  $\pm$   0.9 &    2.2 $\pm$   0.4 &    1.6 $\pm$   0.1 &   14.8  $\pm$   0.5 \\
157.43 &   6.19 &    73 &    1.3  $\pm$   0.3 &    4.6  $\pm$   1.0 &    5.3 $\pm$    0.9 &    0.4 $\pm$    0.1 &    2.7 $\pm$    0.5 &    3.1  $\pm$   1.0 &    0.7 $\pm$    0.4 & $<$4.8 &    3.5 $\pm$   0.7 &    2.0  $\pm$   0.5 &   11.7 $\pm$    0.3 \\
166.63 &  -2.16 &   111 &    3.9 $\pm$    0.2 &    9.5 $\pm$    0.7 &    9.1 $\pm$    0.5 &    0.8 $\pm$   0.2 &    4.2  $\pm$   0.4 &   11.1  $\pm$   0.5 &    2.2  $\pm$   0.3 &  $3.6$  $\pm$  1.3 &    2.7 $\pm$    0.4 &    1.5 $\pm$    0.2 &   14.1 $\pm$    0.4 \\
166.63 &   2.01 &   111 &    6.3 $\pm$    0.5 &   15.3 $\pm$    1.3 &   13.5 $\pm$    1.0 &    2.3 $\pm$    0.2 &    6.5  $\pm$   0.6 &   19.4  $\pm$   1.2 &    6.7 $\pm$    0.5 &  22.6   $\pm$  2.1 &    2.1 $\pm$   0.5 &    1.2   $\pm$  0.1 &   16.2   $\pm$  0.6 \\
166.63 &   4.10 &    90 &    4.0 $\pm$    0.4 &    9.4 $\pm$    1.0 &    7.6  $\pm$   0.7 &    2.5 $\pm$    0.2 &    3.2  $\pm$   0.5 &   11.3 $\pm$    0.9 &    6.6 $\pm$    0.6 & 12.3 $\pm$    2.1 &    1.8  $\pm$   0.6 &    1.3  $\pm$  0.2 &   19.2 $\pm$   0.9 \\
169.70 & -12.61 &   110 &    2.0  $\pm$   0.6 &    6.6 $\pm$    2.3 &    7.0  $\pm$   2.4 &    0.5 $\pm$    0.3 &    6.5 $\pm$    1.3 &   14.7  $\pm$   1.9 &    5.4 $\pm$    1.1 &  $8.9$  $\pm$   4.4 &    3.7 $\pm$    0.9 &    1.1  $\pm$   0.4 &   11.8 $\pm$    0.3 \\
178.91 &  -6.34 &    93 &    4.0 $\pm$    0.8 &    6.3 $\pm$    2.3 &    1.1 $\pm$    2.0 &    1.5  $\pm$   0.3 &    2.6 $\pm$    1.5 &   $<$4.8 &  $<$1.2 & -&    8.5 $\pm$     4.5&    3.8 $\pm$     2.0 &    9.3 $\pm$    0.4 \\
178.91 &   6.19 &    88 &    2.7  $\pm$   0.7 &    9.8  $\pm$   1.6 &    9.6 $\pm$    1.1 &    1.3  $\pm$   0.2 &    6.0 $\pm$    0.5 &   19.1  $\pm$   1.3 &    5.7 $\pm$    0.7 & $<$4.9 &    3.0  $\pm$   0.7 &    1.0 $\pm$    0.2 &   12.8 $\pm$    0.4 \\

\hline
Median & values  &&&&&&&&&&&&\\
  - &   - &    - &    3.8 $\pm$    0.8 &   12.0 $\pm$    2.0 &   11.7 $\pm$    1.5 &    2.1 $\pm$    0.2 &    6.5 $\pm$    0.6 &   15.7 $\pm$    1.2 &    4.4 $\pm$    0.6 &  $8.8$ $\pm$    2.0 &    2.7 $\pm$   0.7 &    1.7 $\pm$    0.2 &   14.0 $\pm$    0.5 \\

\hline
\hline
\end{supertabular}
\newpage

\tablefirsthead{\hline \hline \multicolumn{1}{l}{\small Long.} &
\multicolumn{1}{l}{Lat.} & pix. & \multicolumn{1}{l}{b$_{100}$} & \multicolumn{1}{l}{b$_{140}$} & \multicolumn{1}{l}{b$_{240}$} & \multicolumn{1}{l}{b$_{550}$} & \multicolumn{1}{l}{b$_{850}$} & \multicolumn{1}{l}{b$_{1382}$} & \multicolumn{1}{l}{b$_{2098}$} & \multicolumn{1}{l}{b$_{3191}$} & \multicolumn{1}{l}{$\rm \beta_{CO}^{FIR}$} & \multicolumn{1}{l}{$\rm \beta_{CO}^{submm}$} & \multicolumn{1}{l}{T$\rm _d^{CO}$} \\ 
 &&&\multicolumn{1}{l}{$\pm\Delta$b$_{100}$}&\multicolumn{1}{l}{$\pm\Delta$b$_{140}$}&\multicolumn{1}{l}{$\pm \Delta$b$_{240}$}&\multicolumn{1}{l}{$\pm \Delta$b$_{550}$}&\multicolumn{1}{l}{$ \pm \Delta$b$_{850}$} &\multicolumn{1}{l}{$\pm \Delta$b$_{1382}$} & \multicolumn{1}{l}{$\pm \Delta$b$_{2098}$} & \multicolumn{1}{l}{$ \pm \Delta$b$_{3191}$} & \multicolumn{1}{l}{$\rm  \pm \Delta \beta_{CO}^{FIR}$} &\multicolumn{1}{l}{$\rm  \pm \Delta \beta_{CO}^{submm}$} & \multicolumn{1}{l}{$\pm \Delta $T$\rm _d^{CO}$} \\
& & & & && & ($\times 10^{-1}$) &  ($\times 10^{-2}$) &  ($\times 10^{-2}$) &  ($\times 10^{-3}$) & & & \\\hline}
\tablehead{\hline \multicolumn{14}{l}{\small\sl continued from previous page}
\\ \hline \multicolumn{1}{l}{\small Long.} &
\multicolumn{1}{l}{Lat.} & pix. & \multicolumn{1}{l}{b$_{100}$} & \multicolumn{1}{l}{b$_{140}$} & \multicolumn{1}{l}{b$_{240}$} & \multicolumn{1}{l}{b$_{550}$} & \multicolumn{1}{l}{b$_{850}$} & \multicolumn{1}{l}{b$_{1382}$} & \multicolumn{1}{l}{b$_{2098}$} & \multicolumn{1}{l}{b$_{3191}$} & \multicolumn{1}{l}{$\rm \beta_{CO}^{FIR}$} & \multicolumn{1}{l}{$\rm \beta_{CO}^{submm}$} & \multicolumn{1}{l}{T$\rm _d^{CO}$} \\ 
 &&&\multicolumn{1}{l}{$\pm\Delta$b$_{100}$}&\multicolumn{1}{l}{$\pm\Delta$b$_{140}$}&\multicolumn{1}{l}{$\pm \Delta$b$_{240}$}&\multicolumn{1}{l}{$\pm \Delta$b$_{550}$}&\multicolumn{1}{l}{$ \pm \Delta$b$_{850}$} &\multicolumn{1}{l}{$\pm \Delta$b$_{1382}$} & \multicolumn{1}{l}{$\pm \Delta$b$_{2098}$} & \multicolumn{1}{l}{$ \pm \Delta$b$_{3191}$} & \multicolumn{1}{l}{$\rm  \pm \Delta \beta_{CO}^{FIR}$} &\multicolumn{1}{l}{$\rm  \pm \Delta \beta_{CO}^{submm}$} & \multicolumn{1}{l}{$\pm \Delta $T$\rm _d^{CO}$} \\
 & & & && & & ($\times 10^{-1}$) &  ($\times 10^{-2}$) &  ($\times 10^{-2}$) &  ($\times 10^{-3}$) & & & \\  \hline}
\tabletail{\hline\multicolumn{14}{r}{\small\sl continued on next page}\\\hline } 
\tablelasttail{\hline}
\bottomcaption{Results of the IR/CO correlations in the molecular phase for all sky regions.
Cols 1-2: central position of each studied regions in degree. Col 3: number of pixels used for the correlations in each studied regions. Cols 4-11: results of the correlations in $\rm {MJy/sr}$ for $\rm 10^{21}H/cm^{2}$. Cols 12-13: emissivity
spectral indices in the range 100 - 240 $\rm \mic$ ($\rm \beta^{FIR}$)
and in the range 550 - 2100 $\rm \mic$ ($\rm \beta^{submm}$). Col 14: derived dust temperature in Kelvin. Errors bars represent the 1-$\rm \sigma$ uncertainty estimates. Uncertainties equal to 0.0 are inferior to 0.05.
}
\begin{supertabular}{p{0.95cm}p{0.91cm}p{0.3cm}p{1.6cm}p{1.6cm}p{1.6cm}p{1.5cm}p{1.5cm}p{1.5cm}p{1.6cm}p{1.6cm}p{1.32cm}p{1.3cm}p{1.5cm}}
\hline
\hline
 Case 1  &&&&&&&&&&&&&\\
\hline
 08.34 &   8.28 &    71 &    8.8 $\pm$    1.1 &   18.6  $\pm$   2.0 &   13.9  $\pm$   1.2 &    1.8  $\pm$   0.2 &    6.0 $\pm$    0.3 &   12.5 $\pm$    0.7 &    3.0 $\pm$    0.4 &  $9.2$  $\pm$   1.6 &    2.4 $\pm$   0.5 &    1.7  $\pm$   0.2 &   16.7 $\pm$    0.6 \\
117.55 &  -2.16 &   111 &    5.0  $\pm$   0.9 &   11.3  $\pm$   1.8 &    9.1 $\pm$    1.2 &    1.2 $\pm$    0.2 &    3.4 $\pm$    0.5 &   11.3  $\pm$   0.8 &    3.3 $\pm$    0.3 &  10.0 $\pm$    1.3 &    2.3 $\pm$    0.6 &    1.2  $\pm$   0.2 &   16.2 $\pm$   0.6 \\
120.61 &  -6.34 &    40 &    9.2  $\pm$   6.2 &   18.2 $\pm$    7.9 &   15.4 $\pm$    4.1 &    1.9 $\pm$    0.4 &    6.5 $\pm$    0.9 &   16.8  $\pm$   2.7 &    5.0 $\pm$    1.3 &  14.0 $\pm$    5.9 &    2.2  $\pm$   1.4 &    1.3  $\pm$   0.3 &   16.8  $\pm$   0.6 \\
139.02 &  12.46 &    17 &    2.8  $\pm$   1.0 &    7.9  $\pm$   2.5 &    7.5  $\pm$   2.9 &    1.3 $\pm$    0.3 &    0.2 $\pm$    1.9 &    6.0  $\pm$   2.4 &    4.0  $\pm$   1.2 &  14.7 $\pm$    7.7 &    2.2  $\pm$  1.1 &    1.4  $\pm$   0.4 &   15.1 $\pm$    0.5 \\
142.09 &   2.01 &   111 &    5.5  $\pm$   0.8 &   14.3  $\pm$   1.3 &   13.0  $\pm$   0.8 &    1.9 $\pm$    0.1 &    5.6 $\pm$    0.3 &   13.1  $\pm$   0.5 &    3.5  $\pm$   0.2 &  $9.4$ $\pm$    0.8 &    2.3 $\pm$    0.5 &    1.6 $\pm$    0.1 &   15.1 $\pm$    0.5 \\
145.16 &  -4.25 &    91 &    1.8 $\pm$    0.3 &    5.1 $\pm$    0.9 &    6.2  $\pm$   0.8 &    1.1$\pm$     0.3 &    4.1 $\pm$    0.6 &    8.0  $\pm$   0.9 &    2.3 $\pm$    0.6 & $<$4.7 &    1.8 $\pm$    0.6 &    1.6 $\pm$    0.3 &   15.3   $\pm$  0.5 \\
148.23 &  -6.34 &    38 &    2.6  $\pm$   0.5 &   10.3 $\pm$    1.6 &   11.8 $\pm$    1.2 &    0.5  $\pm$   0.2 &    6.4 $\pm$    1.0 &   20.0  $\pm$   1.7 &    6.6 $\pm$    0.8 &  $8.0$ $\pm$    2.9 &    4.8  $\pm$   0.6 &    0.9  $\pm$   0.2 &   10.3 $\pm$    0.2 \\
148.23 &   2.01 &   111 &    3.0  $\pm$   0.3 &    9.6 $\pm$    0.7 &    9.8 $\pm$    0.5 &    1.6  $\pm$   0.1 &    5.4 $\pm$    0.4 &   11.7  $\pm$   0.6 &    3.1  $\pm$   0.3 &  $6.7$ $\pm$    0.8 &    2.4  $\pm$   0.5 &    1.6 $\pm$    0.1 &   14.1  $\pm$   0.4 \\
151.29 &  -0.07 &   110 &    5.3  $\pm$   0.8 &   10.9 $\pm$    1.3 &    9.6  $\pm$   0.8 &    1.4 $\pm$    0.1 &    3.0 $\pm$    0.4 &    9.1  $\pm$   0.7 &    2.4 $\pm$    0.4 &  $6.5$ $\pm$    1.2 &    2.1 $\pm$    0.5 &    1.6 $\pm$    0.2 &   17.3 $\pm$    0.6 \\
151.29 &   6.19 &   100 &    2.1 $\pm$    0.1 &    5.9 $\pm$    0.3 &    5.7  $\pm$   0.2 &    1.2 $\pm$    0.1 &    2.4  $\pm$   0.2 &    7.5 $\pm$    0.2 &    1.1  $\pm$   0.1 &  $7.4$ $\pm$    0.8 &    2.0 $\pm$    0.4 &    1.8 $\pm$    0.1 &   16.0    $\pm$ 0.6 \\
154.36 & -10.52 &    51 &    1.2 $\pm$    0.1 &    3.9 $\pm$    0.2 &    4.4  $\pm$   0.2 &    0.6 $\pm$    0.0 &    2.3 $\pm$    0.1 &    8.1 $\pm$    0.3 &    2.2  $\pm$   0.1 &  $7.3$ $\pm$    0.6 &    2.4 $\pm$   0.4 &    1.1   $\pm$  0.1 &   13.8 $\pm$    0.4 \\
157.43 &  12.46 &    33 &    2.4 $\pm$    0.4 &    7.7 $\pm$    1.5 &    8.4 $\pm$    1.2 &    1.4  $\pm$   0.4 &    3.2  $\pm$   0.8 &    6.5  $\pm$   1.4 &    0.4 $\pm$    0.9 & $<$13.3 &    2.3 $\pm$    0.6   &  2.1  $\pm$   0.5 &   13.9 $\pm$    0.4 \\
160.50 &  -8.43 &   112 &    1.4 $\pm$    0.2 &    4.5 $\pm$    0.5 &    5.8 $\pm$    0.5 &    0.6 $\pm$    0.1 &    2.8 $\pm$    0.3 &    7.3 $\pm$    0.6 &    1.7  $\pm$   0.3 &  $6.9$ $\pm$    1.2 &    2.9 $\pm$    0.5 &    1.3 $\pm$   0.2 &   12.5 $\pm$    0.3 \\
160.50 &   2.01 &   117 &    2.6 $\pm$    0.2 &    7.0 $\pm$    0.5 &    7.8 $\pm$    0.5 &    1.1 $\pm$    0.1 &    3.1  $\pm$   0.5 &    7.7  $\pm$   0.8 &    2.1  $\pm$   0.3 &  $5.1$ $\pm$    1.6 &    2.1 $\pm$    0.4 &    1.6 $\pm$    0.2 &   14.9  $\pm$   0.4 \\
163.57 &   8.28 &    42 &    1.1 $\pm$    0.3 &    2.9 $\pm$    0.9 &    4.2  $\pm$   0.8 &    0.3 $\pm$    0.1 & $<1.2$ &   $<$3.1 &  $<$0.2 & - &    2.8 $\pm$    0.7 & - &   12.7 $\pm$    0.3 \\
163.57 &  10.37 &    59 &    0.9 $\pm$    0.3 &    2.9 $\pm$    0.9 &    3.5 $\pm$    0.7 &    0.2 $\pm$    0.1 &$<$1.4&  $<3.3$ &  $<0.4$ & - &    3.8 $\pm$    0.7 & - &   11.4 $\pm$    0.3 \\
166.63 &   8.28 &    34 &    1.4 $\pm$    0.3 &    3.8 $\pm$    1.0 &    4.6 $\pm$    0.9 &    0.4 $\pm$    0.1 &    0.3 $\pm$    0.5 &    1.2  $\pm$   1.2 & $<$1.2 & -&    2.6 $\pm$   0.7 & - &   13.3  $\pm$   0.3 \\
172.77 & -10.52 &   110 &    1.3 $\pm$    0.1 &    4.5 $\pm$    0.4 &    5.6 $\pm$    0.4 &    0.7  $\pm$   0.1 &    2.6$\pm$     0.3 &    7.8 $\pm$    0.4 &    1.6  $\pm$   0.2 &  $5.9$ $\pm$    0.9 &    2.7 $\pm$    0.5 &    1.4 $\pm$    0.2 &   12.7  $\pm$   0.3 \\

\hline
Median & values &&&&&&&&&&&&\\
- &   - &    - &    2.6  $\pm$   0.3 &    7.7 $\pm$  1.0 &    7.8  $\pm$   0.8 &    1.2  $\pm$   0.1 &    3.1  $\pm$   0.5 &    8.0  $\pm$   0.8 &    2.3 $\pm$    0.4 &  $6.9$  $\pm$   1.6 &    2.4 $\pm$    0.6 &    1.6  $\pm$   0.2 &   14.9 $\pm$    0.4 \\

\hline
\hline
 Case 2  &&&&&&&&&&&&&\\
\hline
71.53 &   4.10 &    64 &   29.4 $\pm$    2.3 &   57.5  $\pm$   4.7 &   34.9  $\pm$   3.2 &    4.8 $\pm$ 0.3 & $<1.6$ &   24.7 $\pm$   1.5 &    7.1  $\pm$   0.8 &  0.3  $\pm$   3.2 &    2.6   $\pm$  0.5 &   1.8  $\pm$   0.2 &   17.7 $\pm$    0.7 \\
 74.59 &   2.01 &    86 &   12.4  $\pm$  3.2 &   29.1  $\pm$   4.0 &   22.0  $\pm$   1.7 &    2.6 $\pm$    0.2 &    4.8  $\pm$   0.8 &   18.2 $\pm$    0.9 &    4.8 $\pm$    0.4 &  $9.4$ $\pm$    1.1 &    2.6 $\pm$    0.6 &    1.6  $\pm$   0.1 &   15.4  $\pm$   0.5 \\
 77.66 &   4.10 &   104 &   41.6 $\pm$    3.3 &   63.4 $\pm$    4.0 &   34.8 $\pm$    1.7 &    2.6 $\pm$    0.2 &   11.9  $\pm$   0.4 &   24.8 $\pm$    0.8 &    8.6 $\pm$    0.4 &  $7.3$ $\pm$    1.4 &    2.9 $\pm$    0.4 &    1.2 $\pm$    0.1 &   17.5  $\pm$   0.6 \\
 80.73 &   8.28 &    56 &   28.2  $\pm$   5.0 &   56.6 $\pm$    7.4 &   34.9  $\pm$   3.8 &    3.4 $\pm$    0.3 &    2.9  $\pm$   1.2 &   18.1 $\pm$    2.0 &    4.9  $\pm$   1.6 &  25.1 $\pm$    7.7 &    2.9   $\pm$  0.6 &    1.9$\pm$     0.2 &   16.0$\pm$     0.6 \\
 83.80 &  -4.25 &    59 &    7.3 $\pm$    1.4 &   14.4  $\pm$   2.5 &   11.2 $\pm$    1.7 &    1.1 $\pm$    0.2 &    1.3  $\pm$   0.9 &   13.9 $\pm$    1.5 &    3.2  $\pm$   0.6 &  $1.9$ $\pm$    2.3 &    2.5  $\pm$   0.6 &    1.2 $\pm$    0.3 &   16.2  $\pm$   0.5 \\
 89.93 &   2.01 &   111 &    2.1  $\pm$   0.3 &    5.5 $\pm$    0.5 &    5.7  $\pm$   0.3 &    0.8 $\pm$    0.0 &    3.0  $\pm$   0.1 &    7.4 $\pm$    0.2 &    1.7 $\pm$    0.1 &  $4.8$ $\pm$    0.3 &    2.2  $\pm$   0.5 &    1.5 $\pm$    0.1 &   15.0  $\pm$   0.5 \\
 93.00 &  -0.07 &   111 &    5.0 $\pm$    0.8 &   11.8  $\pm$   1.3 &   10.0 $\pm$    0.9 &    1.3 $\pm$    0.1 &    3.8  $\pm$   0.4 &    8.5 $\pm$    0.7 &    1.8 $\pm$    0.3 &  $1.8$ $\pm$    1.1 &    2.4 $\pm$    0.5 &    1.8 $\pm$    0.2 &   15.6 $\pm$    0.5 \\
 99.14 &  -4.25 &    91 &    7.6  $\pm$   1.1 &   19.5  $\pm$   2.6 &   17.4  $\pm$   2.1 &    0.7 $\pm$    0.7 &    0.9  $\pm$   2.0 &   20.4 $\pm$    2.6 &  $<$4.8 & $<$12.5 &    4.0  $\pm$   0.5 & - &   11.9 $\pm$    0.3 \\
 99.14 &  -0.07 &   110 &    3.3$\pm$     0.9 &    7.2 $\pm$    1.3 &    5.9  $\pm$   1.0 &    0.4 $\pm$    0.2 &    1.1 $\pm$    0.4 &    3.1  $\pm$   0.8 &    0.9  $\pm$   0.4 &  $4.8$  $\pm$   1.8 &    3.1 $\pm$    0.7 &    1.4  $\pm$   0.5 &   13.8 $\pm$    0.4 \\
 99.14 &  12.46 &    79 &    1.6 $\pm$    0.4 &    5.7 $\pm$    0.8 &    6.0  $\pm$   0.7 &    0.9 $\pm$    0.1 &    2.8 $\pm$    0.4 &    7.8 $\pm$    0.9 &    1.3  $\pm$   0.6 &  $9.6$  $\pm$   3.6 &    2.6  $\pm$   0.6 &    1.6   $\pm$  0.3 &   13.2  $\pm$   0.4 \\
102.21 &  10.37 &    99 &    2.0  $\pm$   0.4 &    6.2  $\pm$   0.9 &    6.3  $\pm$   0.7 &    0.6 $\pm$    0.1 &    2.9  $\pm$   0.3 &    7.0 $\pm$    0.5 &    1.0 $\pm$    0.3 &  $1.9$ $\pm$    1.0 &    2.9  $\pm$   0.6 &    1.5 $\pm$   0.2 &   13.0  $\pm$   0.4 \\
102.21 &  12.46 &    68 &    1.8  $\pm$   0.3 &    5.8  $\pm$   0.7 &    5.9  $\pm$   0.5 &    0.6 $\pm$    0.1 &    3.1  $\pm$   0.3 &    6.4 $\pm$    0.4 &    0.3 $\pm$    0.3 & $<$3.8 &   3.0 $\pm$     0.5 &    1.5 $\pm$    0.2 &   12.8 $\pm$    0.4 \\
108.34 &  -4.25 &    90 &    2.6 $\pm$    0.3 &    7.9  $\pm$   0.6 &    8.1 $\pm$    0.5 &    1.4 $\pm$    0.1 &    3.8  $\pm$   0.2 &    8.9 $\pm$    0.3 &    1.7  $\pm$   0.1 &  $9.0$  $\pm$   0.8 &    2.2 $\pm$     0.5 &    1.9  $\pm$    0.1 &   14.6 $\pm$     0.5 \\
108.34 &   2.01 &   111 &   15.7  $\pm$   1.1 &   24.2  $\pm$   1.4 &   14.3 $\pm$    0.8 &    1.1 $\pm$    0.1 &    4.4  $\pm$   0.2 &    9.4 $\pm$    0.5 &    2.0 $\pm$    0.2 &  $1.8$  $\pm$   0.4 &    2.8 $\pm$     0.4 &    1.7 $\pm$     0.1 &   17.7  $\pm$    0.6 \\
111.41 &   6.19 &    83 &    3.4 $\pm$    1.6 &    8.5  $\pm$   2.7 &    7.5  $\pm$   1.5 &    0.8 $\pm$    0.2 &    3.9 $\pm$    0.4 &    8.6 $\pm$    0.7 &    1.8 $\pm$    0.3 &  $4.8$  $\pm$   0.8 &    2.7  $\pm$    0.9 &    1.7 $\pm$     0.2 &   14.3 $\pm$     0.4 \\
114.48 &  -0.07 &   110 &   18.9 $\pm$    1.7 &   31.0  $\pm$   2.5 &   19.5 $\pm$    1.4 &    0.8  $\pm$   0.2 &    3.9  $\pm$   0.4 &    8.0 $\pm$    0.7 &    1.9 $\pm$   0.2 &  $8.0$ $\pm$    1.1 &    3.4  $\pm$    0.4 &    1.6 $\pm$     0.2 &   14.6   $\pm$   0.4 \\
117.55 &  -0.07 &   110 &    1.5  $\pm$   1.3 &    6.1  $\pm$   2.6 &    6.2 $\pm$    1.7 &    1.0 $\pm$    0.3 &    2.3 $\pm$    0.6 &    7.3 $\pm$    0.9 &    2.1 $\pm$   0.2 &  $2.4$ $\pm$    1.3 &    3.2 $\pm$     1.7 &    1.4 $\pm$     0.3 &   12.5  $\pm$    0.4 \\
123.68 &   2.01 &   111 &    1.5 $\pm$    0.2 &    5.2 $\pm$    0.4 &    6.1 $\pm$    0.3 &    0.8  $\pm$   0.1 &    2.9 $\pm$    0.1 &    7.3 $\pm$    0.3 &    1.5 $\pm$    0.1 &  $5.1$ $\pm$    0.6 &    2.7  $\pm$    0.5 &    1.6  $\pm$    0.1 &   13.0 $\pm$     0.4 \\
129.82 &   2.01 &   110 &    3.5 $\pm$    0.5 &   10.8 $\pm$    0.9 &   10.3  $\pm$   0.6 &    1.3 $\pm$    0.1 &    5.7  $\pm$   0.3 &   13.0  $\pm$   0.6 &    2.4 $\pm$   0.3 & $<$4.3 &    2.7  $\pm$    0.5  &   1.6 $\pm$     0.1 &   13.6 $\pm$     0.4 \\
135.95 &  -4.25 &    49 &    1.2  $\pm$   0.5 &    5.1  $\pm$   1.1 &    5.4 $\pm$    1.0 &    1.3 $\pm$    0.2 &    3.1 $\pm$    0.5 &   10.3 $\pm$    1.1 &    3.5 $\pm$    0.7 &  19.7 $\pm$   4.0 &    2.7 $\pm$    0.9 &    1.4 $\pm$     0.2 &   12.9 $\pm$     0.4 \\
135.95 &  -2.16 &   106 &    5.1 $\pm$    0.7 &   11.0  $\pm$   1.3 &    8.6 $\pm$    0.9 &    0.7 $\pm$    0.1 &    3.9 $\pm$    0.5 &    8.0 $\pm$    0.8 &    1.3 $\pm$    0.4 &  $5.7$ $\pm$    1.4 &    2.9  $\pm$    0.5 &    1.6  $\pm$    0.3 &   14.6$\pm$      0.4 \\
139.02 &  -4.25 &    62 &    1.8  $\pm$   0.3 &    6.6  $\pm$   0.7 &    7.2 $\pm$    0.6 &    1.2 $\pm$    0.1 &    4.0  $\pm$   0.3 &   11.0  $\pm$   0.7 &    3.2 $\pm$    0.3 &  14.8 $\pm$   1.2 &    2.5 $\pm$     0.5 &    1.3 $\pm$     0.2 &   13.5  $\pm$    0.4 \\
139.02 &  -0.07 &   111 &    8.6 $\pm$    3.0 &   21.0   $\pm$  4.1 &   16.6  $\pm$   2.2 &    2.6 $\pm$    0.3 &    6.7  $\pm$   0.8 &   14.5 $\pm$    1.2 &    3.8 $\pm$   0.4 &  $7.3$ $\pm$    1.2 &    2.4 $\pm$     0.7 &    1.8$\pm$      0.2 &   16.0  $\pm$    0.6 \\
139.02 &   2.01 &   110 &    6.8  $\pm$   2.8 &   17.3  $\pm$   3.8 &   14.8 $\pm$    2.0 &    2.4 $\pm$    0.2 &    5.7 $\pm$    0.6 &   15.8 $\pm$    0.9 &    5.0 $\pm$   0.3 &  14.8 $\pm$   1.3 &    2.3 $\pm$     0.8 &    1.5 $\pm$     0.1 &   15.8$\pm$      0.5 \\
145.16 &   4.10 &   111 &    2.6  $\pm$   0.6 &    6.7  $\pm$   1.0 &    7.0 $\pm$    0.7 &    1.1 $\pm$    0.2 &    2.6 $\pm$    0.3 &    4.9 $\pm$    1.1 &    1.7 $\pm$    0.5 &  $4.8$ $\pm$    3.0 &    2.0 $\pm$     0.6 &    1.9  $\pm$    0.3 &   15.5  $\pm$    0.5 \\
145.16 &  10.37 &   116 &    2.1 $\pm$    0.1 &    6.9  $\pm$   0.4 &    7.3 $\pm$    0.3 &    1.0 $\pm$    0.1 &    3.1 $\pm$    0.2 &    8.5 $\pm$    0.5 &    2.7 $\pm$    0.3 &  $8.2$ $\pm$    1.2 &    2.6 $\pm$     0.4 &    1.3 $\pm$     0.1 &   13.4 $\pm$     0.4 \\
151.29 &   4.10 &   110 &    1.6  $\pm$   0.2 &    5.5  $\pm$   0.4 &    5.7  $\pm$   0.3 &    0.9 $\pm$    0.1 &    2.6 $\pm$    0.2 &    7.9 $\pm$    0.4 &    1.1 $\pm$    0.2 &  $4.2$  $\pm$   0.8 &    2.6$\pm$      0.5 &    1.6  $\pm$    0.1 &   13.5  $\pm$    0.4 \\
154.36 &   4.10 &   111 &    1.8 $\pm$    0.2 &    5.2 $\pm$    0.5 &    5.6  $\pm$   0.4 &    0.8 $\pm$    0.1 &    2.7 $\pm$    0.1 &    7.1  $\pm$   0.4 &    1.7 $\pm$    0.2 &  $2.6$ $\pm$    0.5 &    2.3 $\pm$     0.5 &    1.5$\pm$      0.1 &   14.4 $\pm$     0.4 \\
157.43 &  -0.07 &   110 &    1.3 $\pm$    0.2 &    4.5 $\pm$    0.6 &    5.3  $\pm$   0.6 &    1.3 $\pm$    0.1 &    3.5 $\pm$    0.6 &    6.5 $\pm$    1.2 &    1.6 $\pm$    0.5 &  $5.4$ $\pm$    1.8 &    2.0 $\pm$     0.6 &    2.0 $\pm$     0.2 &   14.5   $\pm$   0.5 \\
160.50 & -20.96 &    72 &    1.4  $\pm$   0.2 &    4.1 $\pm$    0.5 &    4.5 $\pm$    0.3 &    0.5 $\pm$    0.0 &    1.9  $\pm$   0.2 &    5.4 $\pm$    0.3 &    1.2 $\pm$    0.2 &  $4.9$ $\pm$    0.4 &    2.6  $\pm$    0.5 &    1.4  $\pm$    0.2 &   13.5 $\pm$     0.4 \\
160.50 & -18.87 &    91 &    2.2  $\pm$   0.5 &    5.5  $\pm$   1.0 &    5.4 $\pm$    0.6 &    0.4  $\pm$   0.1 &    2.2  $\pm$   0.2 &    6.0  $\pm$   0.5 &    1.1 $\pm$    0.1 &  $4.1$ $\pm$    0.4 &    2.9  $\pm$    0.6 &    1.4   $\pm$   0.2 &   13.5  $\pm$    0.4 \\
160.50 &  -0.07 &   117 &    2.7 $\pm$    0.3 &    8.4 $\pm$    0.8 &    9.3 $\pm$    0.8 &    1.7  $\pm$   0.1 &    3.1  $\pm$   0.6 &    6.8 $\pm$    1.2 &    2.6 $\pm$    0.5 &  $1.8$  $\pm$   1.5 &    2.1 $\pm$     0.5 &    1.9  $\pm$    0.2 &   14.5 $\pm$     0.5 \\
169.70 &  -6.34 &   110 &    0.6 $\pm$    0.1 &    2.7  $\pm$   0.2 &    3.6 $\pm$    0.2 &    0.3 $\pm$    0.0 &    1.8 $\pm$    0.1 &    5.4 $\pm$    0.2 &    1.2 $\pm$    0.1 &  $6.7$  $\pm$   0.8 &    6.5 $\pm$     1.5 &    1.6  $\pm$    0.2 &    8.4  $\pm$    0.4 \\
172.77 & -18.87 &   111 &    0.9 $\pm$    0.1 &    3.3 $\pm$    0.3 &    4.5 $\pm$    0.3 &    1.2 $\pm$    0.1 &    2.9 $\pm$    0.2 &    7.8 $\pm$    0.2 &    2.5 $\pm$    0.1 &  $4.2$ $\pm$    0.6 &    1.6 $\pm$     0.5 &    1.5 $\pm$     0.1 &   14.7  $\pm$    0.5 \\
172.77 & -14.69 &   110 &    1.5 $\pm$    0.2 &    6.5 $\pm$    0.6 &    9.4  $\pm$   0.8 &    1.2 $\pm$    0.1 &    4.8 $\pm$    0.5 &    9.4 $\pm$    0.6 &    2.3 $\pm$    0.2 &  $8.0$  $\pm$   0.8 &    3.1 $\pm$     0.5 &    1.8 $\pm$     0.2 &   11.6 $\pm$     0.3 \\
172.77 &  -8.43 &    91 &    1.1 $\pm$    0.1 &    4.3 $\pm$    0.3 &    4.9  $\pm$   0.3 &    0.8 $\pm$    0.1 &    2.5 $\pm$    0.3 &    7.6 $\pm$    0.3 &    1.4 $\pm$    0.2 &  $3.9$ $\pm$   1.1 &    2.6 $\pm$     0.5 &    1.5$\pm$     0.2 &   13.1 $\pm$     0.4 \\
172.77 &  -6.34 &    81 &    1.2 $\pm$    0.2 &    4.0 $\pm$    0.7 &    5.1 $\pm$    0.6 &    0.7 $\pm$    0.1 &    2.1  $\pm$   0.4 &    6.0 $\pm$    0.7 &    1.8 $\pm$    0.5 &  $6.8$  $\pm$   2.5 &    2.5  $\pm$    0.6 &    1.4$\pm$      0.2 &   13.0 $\pm$     0.3 \\
175.84 &   4.10 &   109 &    8.1 $\pm$    0.5 &   16.0 $\pm$    0.7 &   12.3 $\pm$    0.5 &    0.3 $\pm$    0.1 &    2.5 $\pm$    0.2 &    6.6 $\pm$    0.4 &    1.9 $\pm$    0.2 &  $7.5$  $\pm$   0.9 &    4.3 $\pm$     0.4 &    1.1  $\pm$    0.2 &   12.0  $\pm$    0.3 \\
178.91 & -23.05 &    51 &    3.9 $\pm$    0.8 &    9.7 $\pm$    2.7 &   12.4  $\pm$   2.2 &    1.3 $\pm$    0.5 &    5.3 $\pm$    0.8 &   12.6 $\pm$    1.9 &    4.4 $\pm$    1.4 &  12.1  $\pm$    7.6 &    2.2  $\pm$    0.6 &    1.3  $\pm$    0.4 &   14.3  $\pm$    0.4 \\
178.91 &   2.01 &   110 &    3.1 $\pm$    0.3 &    7.6 $\pm$    0.7 &    7.6 $\pm$    0.6 &    0.6 $\pm$    0.3 &    4.3 $\pm$    0.5 &    9.3 $\pm$    0.9 &    2.2 $\pm$    0.3 &  $4.8$ $\pm$    0.8 &    2.7 $\pm$     0.5 &    1.6$\pm$      0.2 &   13.7 $\pm$     0.4 \\
181.97 & -12.61 &    66 &    0.3 $\pm$    0.2 &    1.3  $\pm$   0.7 &    2.0 $\pm$    0.5 &    0.4 $\pm$    0.1 &    1.9 $\pm$    0.1 &    4.4 $\pm$   0.5 &    1.4 $\pm$    0.3 &  $4.9$ $\pm$    1.7 &    2.3   $\pm$   2.1 &    1.3   $\pm$   0.2 &   12.5  $\pm$    0.3 \\
181.97 &  -8.43 &    48 &    1.3 $\pm$    0.8 &    5.1 $\pm$    2.3 &    8.6  $\pm$   2.0 &    1.2 $\pm$    0.3 &    4.9 $\pm$    0.7 &    9.2 $\pm$    2.0 &    2.1 $\pm$    0.8 &  $4.7$ $\pm$    3.9 & - &    2.6  $\pm$   1.7 &    6.8   $\pm$   1.7 \\
181.97 &  -2.16 &   111 &    2.1 $\pm$    0.3 &    6.7 $\pm$    0.8 &    8.3 $\pm$    0.7 &    0.5 $\pm$    0.2 &    4.2 $\pm$    0.3 &    9.7  $\pm$   0.6 &    2.6 $\pm$    0.3 &  11.9  $\pm$  1.3 &    3.7   $\pm$   0.5 &    1.5 $\pm$     0.2 &   11.5$\pm$      0.3 \\
181.97 &   2.01 &   111 &    3.8 $\pm$    0.4 &    8.9  $\pm$   0.8 &    7.7 $\pm$    0.6 &    0.6 $\pm$    0.2 &    4.2  $\pm$   0.3 &    7.6  $\pm$   0.6 &    1.4 $\pm$    0.4 &  $5.6$  $\pm$   1.5 &    2.9  $\pm$    0.5 &    1.9  $\pm$    0.3 &   14.0 $\pm$     0.4 \\
185.04 & -10.52 &    48 &    0.2 $\pm$    0.2 &    0.6  $\pm$   0.8 &    2.0 $\pm$    0.6 &  $<0.3$ &    2.0  $\pm$   0.2 &    3.6  $\pm$   0.6 &    1.1 $\pm$    0.3 &  $3.5$  $\pm$   1.8 & - &    2.0  $\pm$    0.3 &    9.8  $\pm$    0.2 \\

 \hline
Median & values  &&&&&&&&&&&&\\
  - &   - &  - &    2.2 $\pm$     0.4 &    6.7 $\pm$     0.8 &    7.5 $\pm$     0.7 &    0.9  $\pm$    0.1 &    3.1 $\pm$     0.4 &    8.0$\pm$      0.7 &    1.8 $\pm$     0.3 &  $4.9$ $\pm$     1.3 &    2.6 $\pm$     0.5 &    1.6 $\pm$     0.2 &   13.7 $\pm$     0.4 \\

\hline
\hline
 Case 3  &&&&&&&&&&&&&\\
\hline
 68.46 &   4.10 &    34 &   35.1 $\pm$    3.4 &   60.2  $\pm$   6.7 &   34.9 $\pm$    4.4 &    3.4 $\pm$   0.5 &    3.4  $\pm$   4.2 &   28.5 $\pm$    2.8 &    6.3 $\pm$    1.8 &  7.7  $\pm$   6.4 &    2.7 $\pm$    0.5 &    1.5 $\pm$    0.3 &   17.5 $\pm$    0.6 \\
 77.66 &  -0.07 &    61 &   27.2  $\pm$   2.9 &   40.3 $\pm$    3.2 &   23.8$\pm$     1.3 &    2.5 $\pm$    0.1 &    8.1 $\pm$    0.7 &   17.8 $\pm$    0.8 &    5.4 $\pm$    0.4 &  $6.5$ $\pm$    0.5 &    2.5  $\pm$   0.4 &    1.5 $\pm$    0.1 &   19.6  $\pm$   0.8 \\
 77.66 &   2.01 &   107 &   26.3  $\pm$   2.9 &   39.4 $\pm$    3.2 &   23.8 $\pm$    1.3 &    2.6 $\pm$    0.1 &    8.7 $\pm$    0.6 &   18.3 $\pm$    0.7 &    5.6 $\pm$    0.4 &  $7.4$  $\pm$   0.6 &    2.5 $\pm$    0.4 &    1.5 $\pm$    0.1 &   19.5  $\pm$   0.8 \\
 83.80 &  -0.07 &   105 &   42.2 $\pm$    3.0 &   62.8  $\pm$   4.1 &   34.5 $\pm$    1.9 &    2.7 $\pm$    0.1 &   10.7 $\pm$    0.5 &   20.9 $\pm$    0.9 &    6.7 $\pm$    0.3 &  10.0 $\pm$     0.8 &    2.9 $\pm$    0.4 &    1.4 $\pm$    0.1 &   18.0 $\pm$   0.7 \\
 93.00 & -12.61 &    17 &    1.9 $\pm$    1.3 &    2.6$\pm$     3.6 &    5.8 $\pm$    2.0 &    0.7 $\pm$    0.2 &    2.9 $\pm$    1.1 &    6.9 $\pm$    1.6 &    2.1 $\pm$    1.2 &  10.6  $\pm$   5.7 &    0.9  $\pm$  1.9 &    1.3  $\pm$  0.5 &   20.7  $\pm$   0.8 \\
 96.07 &  -6.34 &    43 &   11.0 $\pm$    3.1 &   20.7 $\pm$    4.0 &   16.7 $\pm$    1.8 &    1.7 $\pm$    0.2 &    5.0 $\pm$    0.3 &   10.7  $\pm$   0.7 &    1.3 $\pm$    0.5 & $<$5.9 &    2.4 $\pm$    0.6 &   1.8 $\pm$    0.2 &   16.5  $\pm$   0.5 \\
 96.07 &  -4.25 &   100 &    9.5  $\pm$   1.4 &   19.2 $\pm$    2.0 &   15.9 $\pm$    1.0 &    2.2 $\pm$    0.1 &    5.4 $\pm$    0.4 &   11.2 $\pm$    1.0 &    2.1 $\pm$    0.4 &  0.4 $\pm$    1.5 &    2.2  $\pm$   0.5 &    2.0  $\pm$   0.2 &   17.1   $\pm$  0.6 \\
102.21 &  -0.07 &   111 &    5.2  $\pm$   2.6 &   10.8 $\pm$    3.9 &    8.5 $\pm$    2.2 &    1.4 $\pm$    0.2 &    3.6 $\pm$    0.8 &    8.6 $\pm$    1.5 &    2.4 $\pm$    0.6 &  $3.9$ $\pm$    2.7 &    2.2  $\pm$   1.1 &    1.6  $\pm$   0.3 &   17.7  $\pm$   0.7 \\
105.27 &  -0.07 &   111 &    4.5 $\pm$    2.5 &   11.8 $\pm$    3.6 &   10.2 $\pm$    2.0 &    2.6 $\pm$    0.3 &    5.3$\pm$     0.6 &   11.8 $\pm$    1.1 &    2.6 $\pm$    0.4 &  $5.1$ $\pm$    1.6 &    2.0  $\pm$   1.1 &    2.0  $\pm$   0.2 &   17.1 $\pm$    0.7 \\
126.75 &  -2.16 &   108 &    4.4 $\pm$    0.9 &    7.7 $\pm$    1.3 &    6.2$\pm$     0.9 &    0.5 $\pm$    0.1 &    2.9 $\pm$    0.4 &    6.5 $\pm$    0.6 &    0.7 $\pm$    0.3 & $<$1.6 &    2.5$\pm$     0.6 &    1.5 $\pm$    0.3 &   16.3 $\pm$    0.5 \\
142.09 &  -2.16 &   111 &    3.9  $\pm$   1.2 &    8.2$\pm$     2.2 &    5.7  $\pm$   1.5 &    1.4 $\pm$    0.2 &    1.6 $\pm$    0.6 &    3.8 $\pm$    1.1 &    0.3$\pm$     0.5 &  $2.1$  $\pm$   1.8 &    2.0  $\pm$   0.9 &    2.8  $\pm$   0.5 &   19.5  $\pm$   0.9 \\
142.09 &  12.46 &    55 &    2.1 $\pm$    0.2 &    6.8 $\pm$    0.7 &    6.7 $\pm$    0.6 &    0.3 $\pm$   0.1 &    2.0 $\pm$    0.7 &    3.3$\pm$     1.2 &    1.8$\pm$     0.4 &  $8.8$  $\pm$   1.9 &    4.4 $\pm$    0.5 &    1.0 $\pm$    0.4 &   11.0  $\pm$   0.2 \\
145.16 &  -2.16 &   110 &    3.0 $\pm$    0.5 &    6.8 $\pm$    1.0 &    6.3 $\pm$    0.9 &    1.5 $\pm$    0.1 &    3.7 $\pm$   0.5 &    9.2 $\pm$    1.1 &    1.3  $\pm$   0.4 &  $1.1$ $\pm$    1.1 &    1.7  $\pm$   0.6 &    1.9 $\pm$    0.2 &   18.4  $\pm$   0.7 \\
148.23 &  -2.16 &   111 &    6.0  $\pm$   0.6 &   11.0$\pm$     1.0 &    9.5 $\pm$    0.8 &    1.4 $\pm$    0.2 &    5.2 $\pm$    0.5 &   11.5 $\pm$    0.9 &    2.9 $\pm$    0.3 &  $8.1$ $\pm$    1.3 &    2.0  $\pm$   0.5 &    1.5 $\pm$    0.2 &   18.4  $\pm$   0.7 \\
154.36 &  -8.43 &    53 &    0.9 $\pm$    0.1 &    3.2 $\pm$    0.2 &    4.1 $\pm$    0.3 &    0.5 $\pm$    0.0 &    1.9$\pm$     0.2 &    7.2 $\pm$    0.3 &    1.4 $\pm$    0.2 &  $4.4$  $\pm$   0.7 &    2.7 $\pm$    0.5 &    1.2  $\pm$   0.1 &   12.6 $\pm$    0.3 \\
154.36 &  -4.25 &    98 &    1.1 $\pm$    0.2 &    4.0 $\pm$    0.6 &    5.3  $\pm$   0.6 &    1.3 $\pm$    0.2 &    2.7  $\pm$   0.4 &    7.1 $\pm$    0.6 &    1.5 $\pm$    0.4 &  $3.9$  $\pm$   1.8 &    1.9  $\pm$   0.6 &    1.9  $\pm$   0.2 &   14.1 $\pm$    0.4 \\
154.36 &  14.54 &    20 &    2.9 $\pm$    0.4 &   11.6$\pm$     2.0 &    9.9$\pm$     1.6 &    1.9 $\pm$    0.4 &    2.7 $\pm$    1.0 &   11.4$\pm$     1.9 &    2.3 $\pm$    1.8 &  $4.3$  $\pm$   6.2 &    3.1   $\pm$  0.7 &    1.9  $\pm$   0.4 &   12.8  $\pm$   0.5 \\
157.43 &   2.01 &   111 &    2.3 $\pm$    0.2 &    5.4 $\pm$    0.3 &    5.5 $\pm$    0.3 &    1.1 $\pm$    0.0 &    3.5 $\pm$    0.2 &    6.6 $\pm$    0.4 &    1.2 $\pm$    0.2 &  $3.2$  $\pm$   0.5 &    1.8   $\pm$  0.4 &    1.9 $\pm$   0.1 &   17.3 $\pm$    0.6 \\
157.43 &   6.19 &    73 &    2.3 $\pm$    0.1 &    7.0 $\pm$    0.3 &    7.4 $\pm$    0.3 &    1.3 $\pm$    0.0 &    4.8 $\pm$    0.2 &   11.0 $\pm$    0.3 &    3.0 $\pm$    0.1 &  10.1  $\pm$   0.8 &    2.1 $\pm$    0.4 &    1.5 $\pm$    0.1 &   14.7  $\pm$   0.5 \\
166.63 &  -2.16 &   111 &    2.7 $\pm$    0.3 &    6.7$\pm$     0.7 &    7.2 $\pm$    0.6 &    2.0 $\pm$    0.2 &    3.6 $\pm$    0.5 &    7.9  $\pm$   0.6 &    1.2 $\pm$    0.3 & $<$3.5 &    1.5 $\pm$    0.5 &    2.3 $\pm$    0.2 &   18.0 $\pm$    0.7 \\
166.63 &   2.01 &   111 &    7.5 $\pm$    1.2 &   16.0 $\pm$    3.2 &   16.6 $\pm$    2.4 &    3.9 $\pm$    0.5 &    9.9 $\pm$    1.6 &   21.3 $\pm$    2.9 &    7.4 $\pm$    1.2 &  11.9  $\pm$   5.3 &    1.5 $\pm$    0.6 &    1.6 $\pm$    0.2 &   19.0 $\pm$    0.7 \\
166.63 &   4.10 &    90 &    3.0 $\pm$    1.0 &    4.4 $\pm$    2.6 &    8.5 $\pm$   2.0 &    1.9 $\pm$    0.6 &    6.6 $\pm$    1.4 &   17.5 $\pm$    2.5 &    6.4 $\pm$    1.7 &  17.2 $\pm$    5.8 &    0.6  $\pm$   0.9 &    1.1 $\pm$    0.3 &   23.2 $\pm$    1.0 \\
169.70 & -12.61 &   110 &    1.4 $\pm$    0.1 &    5.0 $\pm$    0.3 &    6.2 $\pm$    0.3 &    1.3$\pm$     0.0 &    3.6$\pm$     0.1 &    7.0 $\pm$    0.2 &    1.3  $\pm$   0.1 &  $3.8$ $\pm$    0.5 &    2.0  $\pm$   0.4 &    2.0  $\pm$   0.1 &   14.1  $\pm$   0.4 \\
178.91 &  -6.34 &    93 &    3.4 $\pm$    0.3 &   10.6$\pm$     0.7 &   12.0 $\pm$    0.6 &    1.0 $\pm$    0.1 &    6.6 $\pm$    0.5 &   10.9 $\pm$    0.5 &    1.9 $\pm$    0.3 &  0.7  $\pm$   1.2 &    3.1$\pm$     0.4 &    1.8  $\pm$   0.2 &   12.4 $\pm$    0.3 \\
178.91 &   6.19 &    88 &    2.2 $\pm$    0.3 &    5.3 $\pm$    0.7 &    5.4 $\pm$   0.5 &    0.9 $\pm$    0.1 &    4.6 $\pm$    0.3 &    7.1 $\pm$    0.6 &    0.9  $\pm$   0.3 &  0.3 $\pm$    1.1 &    2.0  $\pm$   0.5 &    1.8$\pm$     0.2 &   16.1 $\pm$    0.5 \\
 \hline
Median & values  &&&&&&&&&&&&\\
- &  - &   - &    3.4   $\pm$  0.9 &    8.2  $\pm$   2.0 &    8.5  $\pm$   1.0 &    1.4$\pm$    0.1 &    3.7  $\pm$   0.5 &   10.7  $\pm$   0.9 &    2.1  $\pm$   0.4 &  $4.3$ $\pm$    1.5 &    2.1 $\pm$   0.5 &    1.6 $\pm$    0.2 &   17.3  $\pm$   0.6 \\

\hline
\hline
\end{supertabular}

\end{landscape}
\newpage
\twocolumn

\end{document}